*Title*

# Environmental impact assessment for climate change policy with the simulation-based integrated assessment model E3ME-FTT-GENIE

*Authors names and affiliations*

**Jean-Francois Mercure**[1,2,3], **Hector Pollitt**[2,3], **Neil R. Edwards**[4], **Philip B. Holden**[4], **Unnada Chewpreecha**[2], **Pablo Salas**[3], **Aileen Lam**[3,5], **Florian Knobloch**[1] and **Jorge E. Vinuales**[3]

[1]Radboud University, Netherlands, Department of Environmental Science, Radboud University, PO Box 9010, 6500 GL, Nijmegen, The Netherlands

[2]Cambridge Econometrics Ltd., Covent Garden, Cambridge CB1 2HT, UK

[3]Cambridge Centre for Environment, Energy and Natural Resource Governance (C-EENRG), University of Cambridge, 19 Silver Street, Cambridge CB3 1EP, UK

[4]Environment, Earth and Ecosystems, The Open University, Milton Keynes, UK

[5]Department of Economics, Faculty of Social Sciences, Humanities and Social Science Building, University of Macao, E21, Avenida da Universidade, Taipa, Macao, China

*Corresponding author*

Jean-Francois Mercure,

Email: J.Mercure@science.ru.nl

Tel: +31 24 36 53256

# Environmental impact assessment for climate change policy with the simulation-based integrated assessment model E3ME-FTT-GENIE

J-F Mercure, H. Pollitt, N.R. Edwards, P.B. Holden, U. Chewpreecha, P. Salas, A. Lam, F. Knobloch, J. E. Vinuales

## Abstract

A high degree of consensus exists in the climate sciences over the role that human interference with the atmosphere is playing in changing the climate. Following the Paris Agreement, a similar consensus exists in the policy community over the urgency of policy solutions to the climate problem. The context for climate policy is thus moving from agenda setting, which has now been mostly established, to impact assessment, in which we identify policy pathways to implement the Paris Agreement. Most integrated assessment models currently used to address the economic and technical feasibility of avoiding climate change are based on engineering perspectives with a normative systems optimisation philosophy, suitable for agenda setting, but unsuitable to assess the socio-economic impacts of a realistic baskets of climate policies. Here, we introduce a fully descriptive, simulation-based integrated assessment model designed specifically to assess policies, formed by the combination of (1) a highly disaggregated macro-econometric simulation of the global economy based on time series regressions (E3ME), (2) a family of bottom-up evolutionary simulations of technology diffusion based on cross-sectional discrete choice models (FTT), and (3) a carbon cycle and atmosphere circulation model of intermediate complexity (GENIE-1). We use this combined model to create a detailed global and sectoral policy map and scenario that sets the economy on a pathway that achieves the goals of the Paris Agreement with >66% probability of not exceeding 2°C of global warming. We propose a blueprint for a new role for integrated assessment models in this upcoming policy assessment context.

## 1. Introduction

### 1.1. New questions raised by the Paris Agreement and the role of models

December 2015 saw a historical moment for climate policy in which, for the first time, almost all countries of the world adopted a formal agreement to reduce emissions in order to limit global warming to temperatures below 2°C [1].[1] This event marked a change in efforts to develop climate policy: the agenda, whether or not to adopt measures to avoid climate change, was mostly set. What remained to be done was to find out how to achieve this objective with public policies, in every country that is party to the agreement.

Developing climate policy is a complex process that could involve planning for dramatic societal changes and socio-economic impacts [2]. Policies can have unintended effects. The far-reaching consequences of adopting particular emissions reduction policies can be challenging to fully foresee, as they involve changes in many sectors and for many actors. For example, could adopting a high price of carbon to incentivise electrification increase electricity prices for consumers, thereby reducing access to modern energy for those who cannot afford it? Can biofuels policy lead to unintended land-use change, or lead to water or food scarcity? Could reducing the consumption of fossil fuels globally lead to high rates of unemployment in producer countries? Could a highly capital-intensive, low-carbon transition lead to excessive debt leveraging of government and/or firms, and result in a carbon bubble?

In order to determine the impacts of specific policies, research must move from the agenda-setting stage to the actual impact assessment of policies. This corresponds to a different stage of the policy cycle, and requires analysing the impacts of detailed baskets of policies, as they are envisaged by policy-makers, with all the attendant political and legal complexities, rather than merely recommending – often unrealistic – policies that appear optimal. In the perspective of impact assessment (e.g. see [3]) the policy parameter space is too large to optimise, and individual

[1] Article 2a of the Paris Agreement states: "Holding the increase in the global average temperature to well below 2°C above pre-industrial levels and pursuing efforts to limit the temperature increase to 1.5°C above pre-industrial levels, recognizing that this would significantly reduce the risks and

policies can synergise or interfere [4]. The complexity of the impact assessment problem must account for the uncertainty over the knowledge of the modeller about the way in which decision-making actually takes place with agents [5], and how the heterogeneity of agents might influence policy outcomes [6, 7]. Models based on representative agents have therefore insufficient resolution for carrying out realistic impact assessment [8]. It is more and more recognised that increasing the level of behavioural information in models enables them to represent more policy instruments and thus cover a wider policy space [9-12].

Climate policy analysis, in the agenda setting perspective (e.g. [13-15]), has focused primarily on total energy system cost, consumption loss and GDP loss as indicators to characterise the socio-economic impacts. This is now insufficient, as policy-makers are increasingly requiring information on many other types of impact [16]. For example, questions arise over large-scale finance of technological change, and its impact on the macroeconomic system [17]. The choice of model type for this purpose pre-determines the results that can be reached [18]. Most equilibrium models of the economy used to analyse climate policy have restrictive assumptions over the functioning of the financial sector such that their outcomes are almost entirely determined by a debatable assumption, that re-allocating finance for technological change to reduce emissions takes away investment from other productive sectors of the economy, which automatically leads to loss of GDP ([19], see also [13] and references therein). In fact, research on innovation tends to suggest the reverse [20-22]. Following the financial crisis of 2008, the key question on the mind of many policy-makers is not how many percentage points of GDP loss climate policy might entail, but rather, whether securing large-scale investment is possible without leading countries to financial instability [23-27].

In this paper, we introduce the new integrated assessment model E3ME-FTT-GENIE1, which is designed to tackle the question of environmental impact assessment with the most realistic policy definition currently available, while enabling policy-makers to explore macro-financial issues that may arise from the introduction of such policy. We first describe the policy context that the model attempts to address, as well as the origin and history of economic thought behind its assumptions. We then describe its components: climatology, non-equilibrium macroeconomics and evolutionary technology modelling. We subsequently provide an example of environmental policy analysis under several socio-economic indicators. We conclude with an outlook for future research in the field of integrated assessment modelling.

## 2. Context: fundamental uncertainty in impact assessment

### 2.1. Pervasive property: fundamental uncertainty means no equilibrium

The modelling approach described in this paper is one of simulation. Each part of the E3ME-FTT-GENIE1 modelling framework attempts to represent real world relationships, in terms of accounting balances, physical interactions and human behaviour. This consistency in approach throughout the suite of linked models is crucial to providing insights that are useful to policy-makers. The results from the model are predictions of outcomes based on empirical behavioural and physical relationships observed in the past and the present.

The starting point of this methodology regarding human behaviour is one of fundamental uncertainty [28, 29]. This premise expresses limitations to knowledge and to the knowable for agents that take part in the economic process. This position runs contrary to the assumptions of perfect knowledge and/or perfect foresight that underlie many other modelling tools, which are used in order to simplify theories and models to a tractable state. Fundamental uncertainty recognises that it is not possible for individuals, firms or other agents to know all the possible outcomes from a decision-making process, and thus that 'unknown unknowns' exist. Under these conditions, it is not possible to estimate probabilities of different outcomes of particular agent decisions, as, with unknown outcomes, the probabilities would never sum to one. From this standpoint, some aspects of decision-making by agents lacking knowledge cannot be reduced to pure risk (as it is in standard Expected Utility Theory). Hence, it is therefore not possible to optimise the decision-making process, and agents either make decision errors, or plan ahead for uncertain outcomes (e.g. with spare production capacity).

As noted by Keen [30], it only requires one agent to make sub-optimal decisions for the system of optimisation to break down as a whole. The consequences are profound. For example, without full knowledge by every economic agent of supply- and demand-price elasticities, there is no guarantee that prices will move to market-clearing rates, where resources would be used in the most efficient manner. The level of output is no longer determined by supply-side constraints (e.g. the number of factories), as the available resources will not necessarily be used (there may be too many factories for the demand). Alternatively, given fundamental uncertainty in the knowledge of the demand function by agents, agents may decide to build spare capacity in preparation for possible demand fluctuations.

### 2.2. There is no optimality in policy-making

Without optimizing behavior, it is not possible to design optimal policy. Probst and Bassi [2] recognize the shortcomings of attempting to optimize public policy. The authors advocate an approach that is based on identifying policy that is found to be effective in the real world, rather than aiming for optimal outcomes. Learning-by-doing in policy-making reduces fundamental uncertainty. To be effective, the policies must first address the issue they are designed for, but ideally, also avoid negative consequences in other policy areas (for example, large economic costs or negative impacts on social cohesion). Due to the complex nature of contemporary economies and the heterogeneous nature of agents that interact within these economies [8], it is not sufficient to monetise these impacts and sum them together using a cost-benefit analysis approach; each must be considered in its own right. Importantly, policies must also be considered in the context of political and legal feasibility (*ibid*). Policy-making does not take place in a political and legal vacuum. The enactment of some policies (e.g. a top-down global carbon price or a standardized income tax rate across countries) may be highly unrealistic and even counterproductive. In some cases, such policies may fall foul of fundamental tenets of social organization enshrined in constitutions or treaties (e.g. human rights provisions) or, due to the limited political space left for their adoption, they may have to be legally structured in a manner that makes them less resilient (e.g. local content requirements in green industrial policy or the use of regulations under scattered statutes [31]).

These findings suggest, for example, that policies based on estimates of the social cost of carbon could be misguided. The assessment approach adopted by the European Commission [3], which follows a method of multi-criteria analysis with extensive stakeholder interaction is much more viable. Under this approach, a limited set of feasible policy options are identified and these are tested across a range of key assessment indicators. This method is applied to all policy proposals, not just those relating to sustainability. This is likely a valid blueprint for successful evidence-based policy-making elsewhere in the world.

### 2.3. Path-dependence and the need for simulation models

Perhaps the key aspect that must be properly accounted for in sustainability transition scenarios, and in macroeconomics in general, is technological and productivity change. Economists here fall into two schools of thought: some consider that technology cannot be influenced by policy and therefore that the economy must adapt to existing 'exogenous' technological change (e.g. robotisation), whereas others think that technology is 'endogenous' and can be influenced by targeted policy. There is extensive empirical evidence that supports the latter position by showing how public policy plays a key role in promoting and guiding technological change [20, 32]. The work of Grubb provides a review of the process of technological development and diffusion in the context of decarbonisation and low-carbon transition [32]. He finds that the rate and direction of technological change can undoubtedly be influenced by policy, and that different types of policy instruments are suitable for different stages of technology development and diffusion. Therefore, a modelling tool that aims to match reality as closely as possible must account for this finding. On the other hand, it is far from demonstrated empirically that the economy can indeed 'internalise' externalities using only pricing incentives, as suggested by standard welfare economics.

Yet, this finding is not new. The work of Arthur showed, using simple models, that relatively minor changes to policy could lead to qualitatively different outcomes for technology diffusion in the long run due to 'social influence', 'path dependency' and technology 'lock-in' [33, 34]. These processes describe how a single technology can come to dominate a particular sector, with highly non-linear outcomes. This is also a key finding in the study of the diffusion of innovations [35]. Policy-makers can steer users towards a particular technology but the rates of technology adoption are, again, highly complex, with considerable uncertainty about the outcomes. Modelling path-dependent systems requires simulation models, since the behaviour of such systems, by definition, depends on relationships between present and past conditions. Optimisation methodologies are not suitable to model path-dependence, since they do not make a clear connection between points in time.

## 3. The E3ME-FTT-GENIE integrated assessment model

### *3.1. Overview of the integrated assessment simulation model*

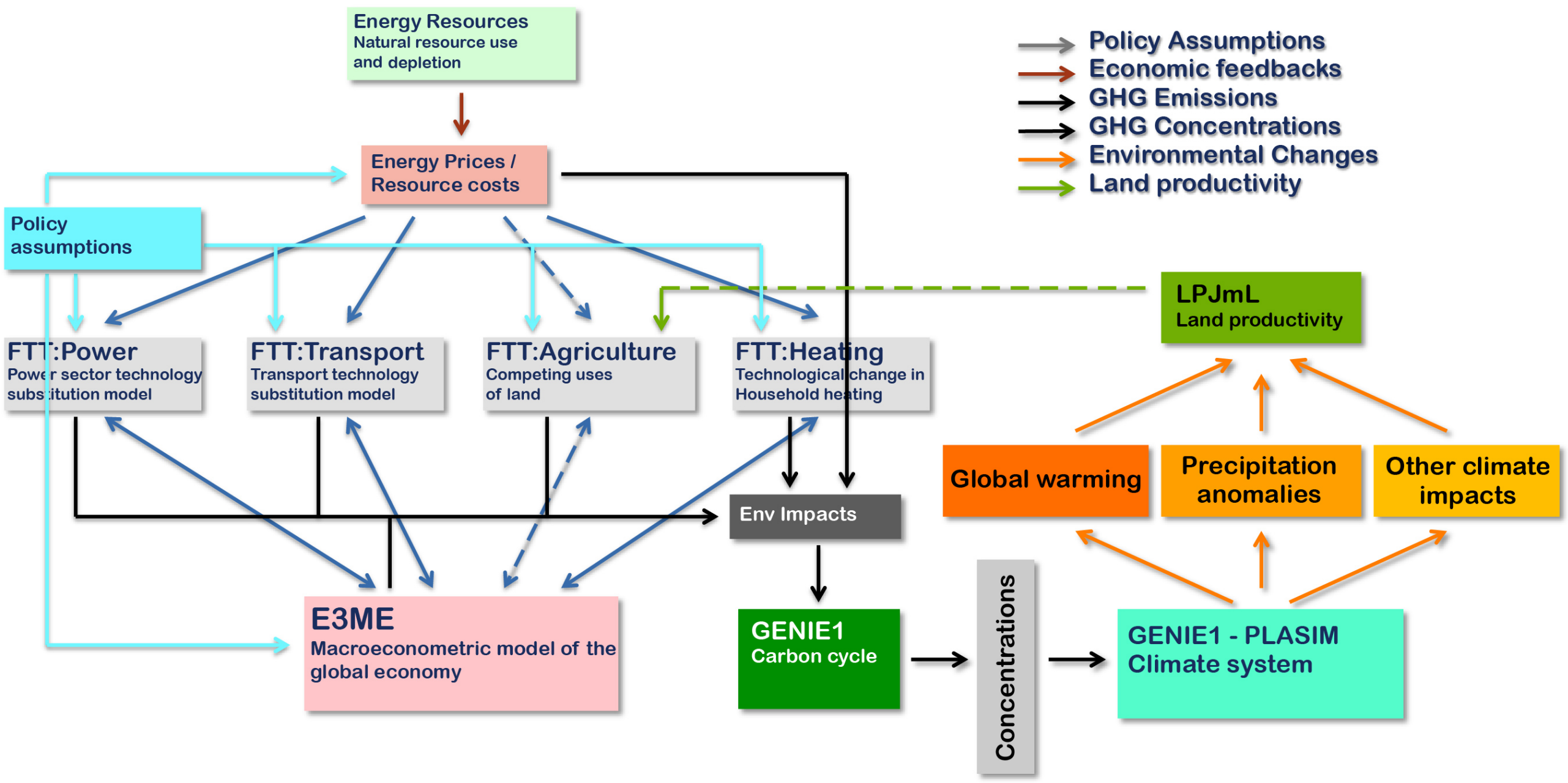

Figure 1: Diagram of the E3ME-FTT-GENIE1 integrated assessment simulation model. Note that the land-use model is under development and not discussed here. Dashed lines refer to sections under development.

The E3ME-FTT-GENIE1[2] model is a simulation-based integrated assessment model that is fully descriptive, in which dynamical (time-dependent) human or natural behaviour is driven by empirically-determined dynamical relationships. At its core is the macroeconomic model E3ME, which represents aggregate human behaviour through a chosen set of econometric relationships that are regressed on the past 45 years of data and are projected 35 years into the future. The macroeconomics in the model determine total demand for manufactured products, services and energy carriers. Meanwhile, technology diffusion in the FTT family of technology modules determines changes in the environmental intensity of economic processes, including changes in amounts of energy required for transport, electricity generation and household heating. Since the development and diffusion of new technologies cannot be well modelled using time-series econometrics, cross-sectional datasets are used to parameterise choice models in FTT. Finally, greenhouse gas emissions are produced by the combustion of fuels and by other industrial processes, which interfere with the climate system. Natural non-renewable energy resources are modelled in detail with a dynamical depletion algorithm. And finally, to determine the climate impacts of chosen policies, E3ME-FTT global emissions are fed to the GENIE1 carbon cycle-climate system model of intermediate complexity. This enables, for instance, policy-makers to determine probabilistically whether or not climate targets are met.

Figure 1 shows the structure of the model. E3ME exchanges information dynamically with several FTT technology diffusion modules, themselves hard-linked to E3ME. E3ME generates the demand for carbon-intensive products and services to the FTT modules, which feed back prices, investment and the demand for other inputs such as energy carriers. The natural resources modules limit the deployment of renewables, and track the depletion of fossil and nuclear fuels. The models are solved together iteratively.

[2] E3ME: Energy-Economy-Environment Macro-Econometric model, global macroeconometric model. FTT: Future Technology Transformations family of technology models. GENIE1: Grid Enabled Integrated Earth system model, an integrated carbon-cycle and climate simulation model.

The model is path-dependent, such that different policy scenarios generate different techno-economic and environmental trajectories that diverge from each other over time. There is no unique parameter (or objective function) under which the model can be optimised in terms of a particular outcome variable, since the space of socio-economic indicators is relatively large, and value judgment is left to the interpretation of the user. Since the sheer size of the policy space is enormous, and the computational task would be relatively intensive, there is likely little value in return for the substantial effort that would be required to optimise policy scenarios. Furthermore, several different types of baskets of policies can reach the same environmental outcomes.

The model is instead used under the 'what if' mode of impact assessment: policies are chosen, and outcomes are observed in terms of the choice of policies. The policies designed in the model are policy instruments that exist in the real world, for example emissions trading schemes, energy taxes, vehicle taxes, feed-in tariffs, subsidies, direct regulation or biofuel mandates. Other assumptions concern government expenditure on education, defence and other services, demography and the price of some global commodities. In the Supplementary Information (SI), we provide a complete cross-referenced list of model equations for E3ME-FTT.

### *3.2. The macroeconomic model E3ME*

E3ME is a computer-based model of the world's economic and energy systems, linked to emissions. It was originally developed through the European Commission's research framework programmes and is now widely used globally for policy assessment, for forecasting and for research purposes. Examples of recent applications include [36-42]. The full manual for the model [43] is available at the model website www.e3me.com. In this section we provide a short summary description. A list of the model's equations is provided in the SI.

E3ME splits the world into 59 global regions, with 43 sectors in each region. The regions are linked through bilateral trade equations, while input-output tables provide the linkages between the different sectors. As a macro-econometric model, E3ME's data requirements are extensive, with time-series data required for each indicator in each sector in each country. The current model database covers 1970-2015 and the main data sources are Eurostat, OECD, World Bank, IMF, IEA and national statistical agencies. The econometric techniques used to specify the functional form of the equations are the concepts of co-integration and error-correction methodology, particularly as promoted by Engle and Granger [44], and Hendry et al. [45] (SI section 1.4). Thus, the model is regressed over the period 1970 to 2015 (45 years), and runs freely between 2016 and 2050 (35 years).

E3ME is often compared to Computable General Equilibrium (CGE) models and is often applied to answer the same sorts of questions, using the same scenario-based approach. The accounting identities described below are in general consistent with the ones that can be found in a CGE model. The inputs to E3ME are also similar to inputs to CGE models and the different modelling approaches share many of the same output indicators.

However, there are key differences between E3ME and a typical CGE model. E3ME is derived from post-Keynesian economic theory [46], as opposed to neoclassical economics for CGE models (See [17, 18]). As stated in section 2.1, the starting point for agents in the model is one of fundamental uncertainty, and behaviour is inferred from past relationships. Although the model is thus subject to the Lucas critique [47], it avoids assumptions about optimisation and perfect information that have been questioned as representations of the real world [30, 48].

Fundamental uncertainty implies that while the identity of supply and demand matching is observed, there is no constraint that demand equals *potential* supply. It is thus possible for there to be unused resources, for example unemployed workers, unused equipment or financial capital, which can be brought in for production if the demand requires it. E3ME incorporates a treatment of endogenous money [46]; the treatment of finance in CGE and other theoretical modelling approaches has been recognised as a major limitation in approach [18], which sits at odds with the observed reality [49, 50].

In such a demand-led economy, it is the level of effective demand that determines output, as originally described by Keynes [28, 51]. The model solves iteratively in the same way as an economic multiplier could be estimated by repeatedly carrying out matrix multiplication of the input-output table (instead of calculating the inverse). But whereas multiplier analysis only calculates changes in intermediate demand, the econometric equations in E3ME make final demand endogenous as well. Crucially, prices are also determined through econometric equations, rather than automatic adjustments that achieve market clearing.

Since E3ME allows for the possibility of the existence of spare resources, it can sometimes yield positive economic and social benefits of technological change policies such as for climate change mitigation, in contrast to CGE models. It is possible (although by no means guaranteed) to predict double dividends[3] in model results, where environmental regulation can lead to faster rates of economic growth [52-54], something that is ruled-out in the premise of CGE models.

E3ME is based on a social accounting matrix that involves highly disaggregated input-output tables. This defines the macroeconomic identity, in which total demand is derived from intermediate demand,

$$L + V = Y = C + I + G + (X - M), \qquad (1)$$

Where *L* represents labour wages, *V* represents remaining value added (profits and taxes on production), *Y* is total production (GDP), *C* is consumption, *I* is investment, *G* is government expenditure, *X* is exports and *M* is imports. Here, *C*, *I* and *M* are derived from econometric regressions, while *Y* is derived from the identity that supply equals total demand (*G* is exogenous). *C* is function of income and prices, while *I* is function of the ratio of actual output to potential output, prices and technological progress, and *M* is function of international competitiveness and technological progress; total imports are split into bilateral flows which, when inverted, can be summed to yield exports. Technological progress is measured by cumulating past investment and R&D, and gives rise to price declines as technology improves; and thus a virtuous cycle arises between prices, consumption, exports, investment and output, the origin of endogenous growth in E3ME.

This contrasts with a CGE model in which *Y* is determined by a production function, *C* is derived through the macroeconomic identity (1), *I* is equated to savings, a fixed proportion of *Y* and technology is often exogenous. In E3ME, *I* is not function of *Y*, but is indirectly related through the fact that when economic growth rates increase, more investment opportunities arise. This is described in detail in the SI.

### 3.3. Technological change model family FTT

The diffusion of individual types of technology (e.g. electric vehicles, wind turbines) is not correctly modelled using linear regression models applied to time series such as those used for each econometric specification in E3ME (e.g. for energy demand). This is because the diffusion of innovations typically follows network effects, where adoptions of or investment in new technologies by agents are strongly influenced by whether other agents have done the same previously (see [55]). This can be due to the fact that agents adopt technologies used in their surroundings with a higher likelihood than technologies they have no experience of, and/or to the fact that firms with higher sales volumes are more able to capture market shares, while innovations with comparatively small sales volumes are produced by firms with lower capacity (expanding capacity takes time and requires expectations of future sales). Both of these processes lead to archetypical S-shaped diffusion curves (see [35, 56]).

In order to represent diffusion, regression models would need to regress a variable onto itself, leading to a recursive endogenous problem that does not have a unique parametric solution [55]. This is a reflection of the fact that diffusion is a path-dependent process: it strongly depends on its past history (e.g. see [33]). This also means that diffusion builds momentum as it progresses, since the faster the diffusion of an innovation is, the faster it can become. In a model, this mathematical property (autocorrelation in time) prevents in fact the model configuration to instantaneously flip between different but simultaneously attractive adoption pathways, making models more intuitive and realistic: once a pathway is followed, it becomes increasingly locked-in. Due to this property, results also become less critically reliant on very detailed cost data, in comparison to the more common social planner/representative agent paradigms of cost-optimisation (i.e. agents do not simply just adopt the very least cost options; they tend to adopt what is already dominant in the market. See [55], [4, 57] for discussions; a clear reason is given in [48]).

[3] Double dividends here mean two simultaneous positive impacts of environmental policy, which on the one hand incentivises agents to address environmental harm, and on the other hand re-allocates funds in a way that improves overall income.

In FTT, we consider agents who own or use a technology that produces a certain service (e.g. generating electricity, transport, household heating), and who consider replacing that technology for a new unit. Such an event takes place at a rate determined by the survival in time of technology units and/or the financing schedule, when switching from *i* to *j*, denoted $A_{ij}$. We assume that these agents make comparisons between options that they see available in the market, which we structure by pair-wise comparisons (other comparison schemes are equivalent, see [55]). The proportion of agents already using technology *i* is $S_i$, that technology's market share. The proportion of agents considering the advantages of technology *j* is $S_j$, the market share of technology *j*. We denote the relative preference of agents for technology *j* over technology *i* with the matrix $F_{ij}$, a fraction between 0 and 1. If we picture shares of technologies being transferred between technology categories as agents gradually replace the stock, then we obtain the equation:

$$\Delta S_i = \sum_{j=1}^{N} S_i S_j \left[ A_{ij} F_{ij} - A_{ji} F_{ji} \right] \Delta t, \qquad (2)$$

This equation is famously named the Lotka-Volterra competition equation, a system of non-linear differential equations more often used in ecology to express the competition for resources between species in an ecosystem (e.g. plants competing for space). It is also extensively used in evolutionary game theory [58]. This equation is used in FTT models under slightly varying types of parameterisation. We describe this mathematical system in detail in [55, 59, 60].

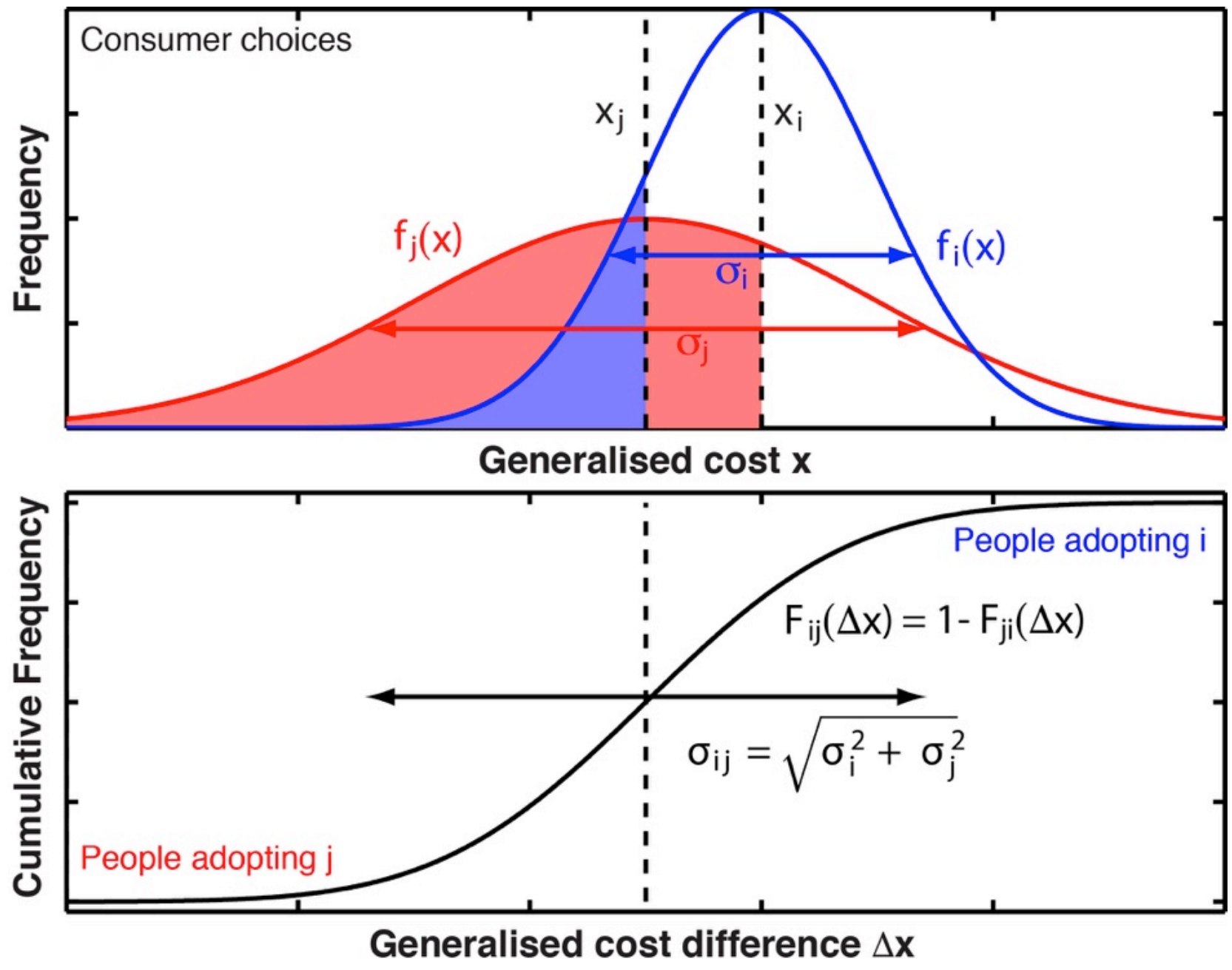

Figure 2 Schematic representation of pair-wise comparison of technological options by heterogenous agents with varying preferences. *(Top panel)* Preferences vary following distributions in an appropriately chosen space of generalised cost. *(Bottom panel)* The resulting choice matrix follows a series of binary logits forming the choice matrix Fij, of which the variations in generalised cost space follows the degree of heterogeneity of agent preferences. Reproduced from [60].

In FTT, we assume that agents minimise their own costs and benefits, but due to multi-agent influence (diffusion networks), this does not generally lead to a cost optimum at the system level, and indeed, we do not optimise total system cost. The preference matrix $F_{ij}$ is probabilistic, determined by the use of a binary logit (fig. 2, see also [61, 62]). Discrete choice theory is used to represent the diversity of agent preferences in a group, a diversity that determines elasticities of substitution.[4] Here, substitution is not instantaneous, as opposed to standard multinomial logit models, due to our use of the Lotka-Volterra dynamical system. Thus we use here a binary logit to determine preferences, not substitutions. Substitutions follow preferences, but also availability. It has, however, the key property of representing heterogeneity of preferences, which translates in the model as a probabilistic nature for $F_{ij}$ (e.g. $F_{ij}$ = 30% and $F_{ji}$ = 70% means that 30% of agents, who have knowledge of both technologies *i* and *j*, prefer technology i, while 70% of these prefer technology *j*).

The binary logit is calculated as follows. We define a generalised cost axis *C* that encompasses all relevant quantifiable components of preferences. The diversity of agents is represented as distributions of perceptions of agents over this cost axis (fig. 2). The comparison exercise becomes one of comparing probability distributions. We assume these distributions to be normal in a space simply functionally related to *C* (linearly in the power sector and heating models, see Mercure 2012 [59], lognormal distributions for transport, see [7, 57]). The result of the comparison of distributions is the binary logit [60]:

$$F_{ij} = \frac{1}{1 + \exp\left(\frac{C_j - C_i}{\sigma_{ji}}\right)}, \quad \sigma_{ji} = \sqrt{{\sigma_i}^2 + {\sigma_j}^2}, \quad F_{ij} + F_{ji} = 1 \qquad (3)$$

This generates the property that cost differences have to be larger than the diversity of the group's perceptions (i.e. $C_j - C_i \gg \sigma_{ji}$) in order for attractiveness towards one option to be significant, and therefore for diffusion to take place. This makes the model less reliant on the accuracy of cost data when diversity is high, which is generally the case ($\sigma_i$ is typically of the order of one third of $C_i$).

Finally, FTT is calibrated on historical diffusion data in order to make its outputs consistent with history. This is done by adding a factor in the cost being minimised by agents, which makes diffusion trajectories at the start of the simulation the same as technological trajectories observed in historical data near to the start date of the simulation (equating the simulated/historical rates of adoption).

More details on the FTT model can be found in [55, 59, 60] (theory) and [4, 57] (computational models). We note that the diffusion of innovations in FTT assumes the deployment of any infrastructure necessary for the deployment of technologies, which we consider part of the diffusion process (see for instance [63]).

### *3.4. A dynamical fossil fuel resource depletion model*

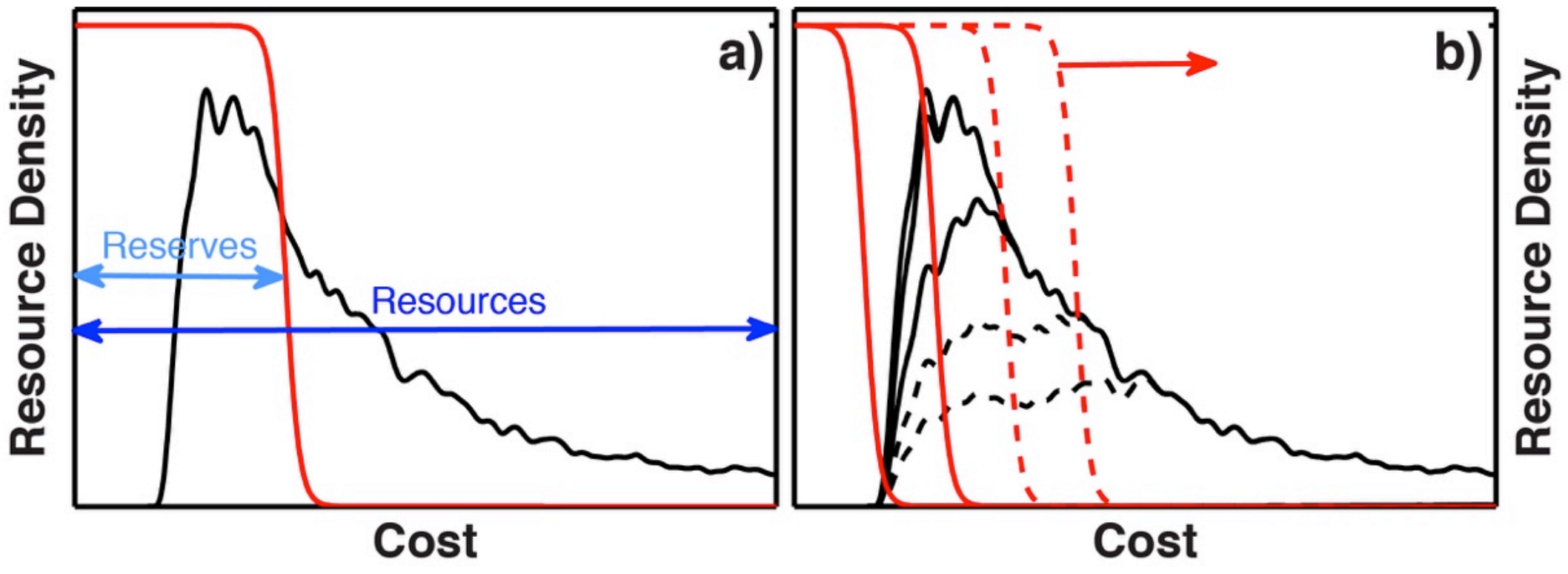

Figure 3: Schematic representation of the non-renewable resource depletion algorithm. a) Resource cost distribution at the start year (black), and extraction likelihood function (red), defining the quantity of reserves. b) With a stable demand, as reserves are gradually extracted and consumed, the reduced extraction rate forces the marginal cost to go up. Reproduced from [64].

The fossil fuel depletion model used in this work is derived from a model by Mercure & Salas [64],

---

[4] Diversity/heterogeneity here means all the sources of variations in decisions between different agents, leading to distributed quantities in the model.

based on a fossil fuel resource dataset given in Mercure & Salas [65]. The model represents global markets for homogenous fossil fuel commodities, which are produced at various locations around the world, with different methods, which incur different costs of production. The main assumption of the model is that the marginal cost (i.e. the cost of the most expensive unit of fuel produced) sets the price of the commodity. This is supported by the assumption that producers refuse to produce at a loss (this is not always strictly true; however, producing at a loss cannot last indefinitely): if the price does not cover producer costs, we assume that producers refuse to supply fuel. Thus, sellers at the marginal cost enjoy low or zero profit over their production, while producers in lower cost ranges enjoy a larger profit. The model does not cover storage for security and price speculation purposes, or supply processing bottlenecks, and therefore does not reproduce some types of cyclic behaviour. The model generates a base price under which lower cost types of production are profitable, and higher cost types of production are unprofitable, the balance of which supplies global demand. Results from this model are used to proportionally scale fuel price changes in E3ME as demand evolves; E3ME prices include taxes and possible margins of profit, not included in the depletion algorithm discussed here.

The depletion algorithm works as follows (fig. 3). The fossil fuel resource database features quantities of fossil fuels at different production costs at the time of the start of the simulation, which were interpolated into a cost distribution of resources. Resources are extracted at a rate determined empirically [64], representing a combination of technical constraints (e.g. oil well technical depletion rates) and human decisions (e.g. strategic choices). We assume the same rate of depletion in all cost ranges based on available information.

We assume that the rate of extraction of resource $i$ in each possible extraction cost ranges $C_i$ to $C_i + dC_i$ is proportional to the quantities left in those cost ranges, with the same proportionality factor $\nu_i$. Thus, if $n_i(C,t)$ is the cost distribution of resources left at time $t$, and $\nu_i$ is the rate of extraction (reserve to production ratios), then the depletion algorithm is:

$$\Delta n_i(C,t)dC_i = \nu_i n_i(C,t) f(P_i - C_i)\Delta t \, dC_i, \qquad (4)$$

where $f(P_i - C_i)$ is a cumulative probability distribution function for the likelihood of deciding to extract resources in cost range C given a commodity price P (a smooth step function equal to zero when the price is much lower than the cost of extraction; equal to one if the extraction is profitable). We denote total supply as $F(t)$, and sum up supply from all types of extraction sites currently operating:

$$F_i(t) = \int_0^\infty \Delta n_i(C,t)dC_i = \int_0^\infty \nu_i n_i(C,t) f(P_i - C_i)\Delta t \, dC_i. \qquad (5)$$

E3ME-FTT provides a total demand for fossil fuels, which depends on their price, due to a combination of elasticities across the model, as well as efficiency changes due to R&D investment, and due to technological change in FTT models (e.g. the diffusion of renewables and electric vehicles). Thus an iterative process is used in order to determine the supply of fuel and price that agrees with both this model and the whole of E3ME-FTT. At each iteration, the price $P_i$ is searched, by a simple trial and error algorithm, such that $F_i$ equates the E3ME-FTT demand, at which point the price is fed to E3ME-FTT, which supplies a new demand value, and the process starts again, until changes in both $F_i$ and $P_i$ are maintained below certain criteria.

### 3.5. Limits to renewables

E3ME-FTT features a database for resource potentials for renewables [65]. This involves, for example, the amount of land suitable for installing wind turbines, the number of potential sites available for hydroelectric installations, or geothermal active areas. These were converted in the form of cost-supply curves. The use of this database is simpler than in the case of non-renewable resources, since all that is necessary is to determine a marginal cost for the resource, which is a function of its level of use. If the level of use increases, the cost increases, while if it goes down (e.g. decommissioning a wind farm), the cost goes down as the sites become available again.

In reality, the process is more complicated than that. The depletion of wind or solar resources takes the form of wind/solar power developers building wind farms or solar parks in areas of low capacity factors, while for hydro or geothermal, it involves higher capital costs if the conditions are less suitable. We thus adjust the appropriate parameters in the calculation of the levelised cost of generating electricity. The depletion of renewable resources leads to higher production costs, which in the binary logit pair-wise comparison, becomes disfavoured over other types of systems, slowing down development. For example, when the good wind power sites have all been developed, investors are faced with using low wind/low capacity factor sites, which for the same wind turbine systems, results in high costs per unit electricity produced, and choose other renewable types instead. The database was developed by the authors using an extensive literature review as well as collected data.

### 3.6. Climate model GENIE

The climate-carbon cycle is simulated with GENIE-1 in the configuration of [66, 67], as applied in the Earth system model of intermediate complexity (EMIC) intercomparison project [68]. GENIE-1 calculates atmospheric $CO_2$ concentrations and climate change from inputs of $CO_2$ emissions, land-use change and non-$CO_2$ climate forcing agents.

GENIE-1 simulates approximately 250 years per CPU hour. This computational speed allows us to provide probabilistic projections, achieved through an 86-member ensemble of simulations for each emissions scenario, varying 28 key model parameters, in order to produce an estimate of the full uncertainty range stemming from uncertainty over these parameters [69]. The computational efficiency of GENIE-1 is achieved mainly through the highly simplified model of the atmosphere, treated as a single layer with horizontal transport that is dominantly diffusive. Computational efficiency also benefits from low spatial resolution (≈ 10° × 5° on average, with 16 depth levels in the ocean) and, relative to high-complexity Earth system models, simplifying assumptions in other model components. These include, for instance, the neglect of momentum transport in the ocean and the representation of all vegetation as a single plant functional type.

The components of GENIE-1 are fully documented in the references that follow. The physical model [70] comprises the 3-D frictional geostrophic ocean model GOLDSTEIN coupled to a 2-D Energy Moisture Balance Atmosphere based on that of Fanning and Weaver [71] and Weaver et al. [72], and a thermodynamic–dynamic sea-ice model based on the work of Semtner [73] and Hibler [74]. Ocean biogeochemistry is modelled with BIOGEM [75], here with phosphate and iron limitation [76, 77] on marine productivity. BIOGEM is coupled to the sediment model SEDGEM [78], describing calcium carbonate preservation in deep-sea sediments and its role in regulating atmospheric $CO_2$. Vegetation and soils are simulated with ENTSML [67], a dynamic model of terrestrial carbon and land use change (LUC), based on the Efficient Numerical Terrestial Scheme [79]. ENTSML takes time-varying fields of LUC as inputs.

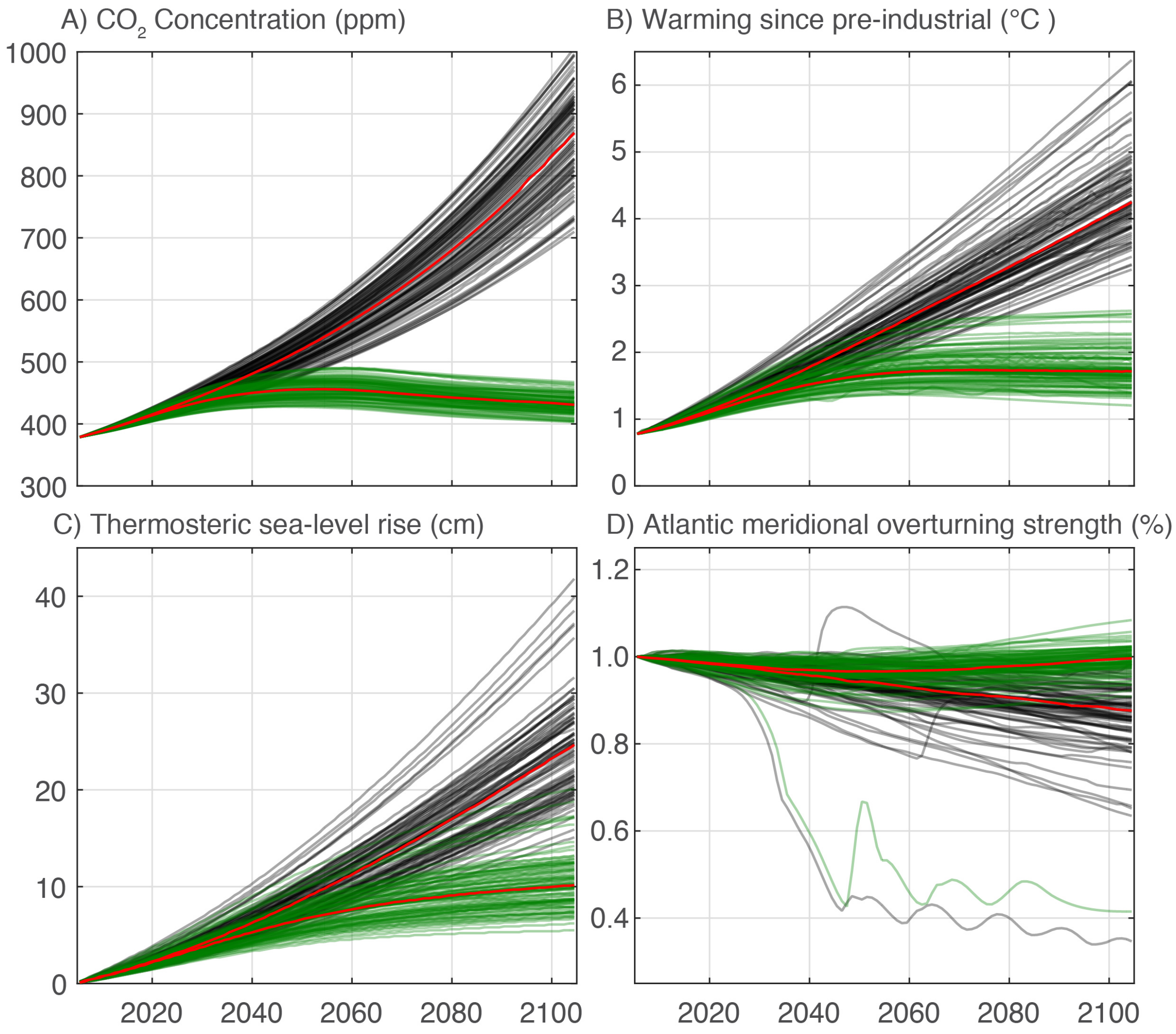

Figure 4: Time evolution for the full 86-member GENIE ensemble of (A) atmospheric CO2 concentration (ppm); (B) warming since preindustrial (°C); (C) global sea-level rise in cm (thermal expansion component); (D) peak Atlantic meridional overturning strength (% of initial). In black are the baseline ensembles, while green are the decarbonisation ensembles. Red lines show median trajectories.

Each GENIE-1 ensemble member continues from an associated simulation in an ensemble of transient historical simulations from AD 850 through to 2005, with forcing described in [80]. Emissions from 2005 are provided by E3ME. Present day $CO_2$ emissions from E3ME are understated by ~1.2GtC, the shortfall arising from neglected processes such as cement production and other small GHG sources. We add 1.2GtC emissions to represent all of these missing sources and apply this adjustment every year scaled by total E3ME emissions.[5] We extrapolate these emissions trajectories until they reach zero post-2050 (see Figure 5 panel i). Other climate forcing agents of non-$CO_2$ trace gases, sulphur emissions and land-use change are taken from an appropriate choice of Representative Concentration Pathway. For instance, RCP2.6 [81] is applied for non-$CO_2$ forcing in strong mitigation scenarios, while RCP8.5 [82] is assumed for a business-as-usual scenario. Different possible extrapolations are given in section 4.2.

More detailed climate impacts (see Figure 4) can be provided by applying the $CO_2$ concentrations output from GENIE-1 to the climate model PLASIM-ENTS or its emulator [83]. We are currently upgrading our capabilities by developing a fully coupled carbon-cycle atmosphere-ocean global climate model, incorporating biogeochemistry into the intermediate complexity AOGCM PLASIM-

[5] Scaling these emissions proportionally to total E3ME emissions can be interpreted as having a policy that aims to phase them out at the same rate as other fuel combustion emissions.

GENIE [84]. The coupling of GENIE to E3ME-FTT is currently made with a soft coupling, from the economy to the climate. The link from climate to the economy through agriculture is under development. This is discussed in the SI. A limitation of this model stems from our current lack of modelling capacity for land-use/land-use change emissions, increasing uncertainty on the result. This will be addressed in an upcoming version of E3ME-FTT that will include land-use modelling, currently under development.

### 3.7. Policy instruments in E3ME-FTT-GENIE1

The goal in the development of E3ME-FTT over time has been to design model representations of policy instruments that resemble real policies as closely as possible. The simulation nature of the model lends itself quite well to that task, as well as the heterogenous agent base of FTT. Policies are of two possible broader types: those applied sector-wide or economy wide, in E3ME, and those applied to specific technologies or applications, in FTT. They are of four possible sub-types: economic incentives (taxes or subsidies), standards/regulations, public procurement and monetary instruments. These are listed in Table 1.

Although the level of resolution afforded by these sub-types is less detailed than what can be analysed through qualitative policy studies or in legal assessments, it is sufficient to capture a diversity of policy instruments as well as their interaction in a way that goes beyond IAMs currently used in climate policy and that is informative for policy-makers. Further refinement of the policy taxonomy is a current area of work in our efforts to improve the modelling approach.

In E3ME, policies are used to influence the behaviour of individuals and firms, as modelled by the econometric equations. For example, a carbon tax influences the amounts of carbon-intensive fuel used in various industries. Policies can also influence investment in particular sectors, and their carbon-intensity by, for example, phasing out the use of coal.

Table 1: Policy instruments in E3ME-FTT

| Policy type | Economy/sector-wide | Technology/process-specific |
|---|---|---|
| Economic incentives | Carbon price, carbon tax, income tax | Technology specific subsidies, taxes, feed-in tariffs (power, vehicles, heating) |
| Standards and regulations | Exogenous* phase-out and efficiency assumptions | Power sector: endogenous* phase-out<br>Road vehicles: efficiency standards, phase out, biofuel mandates<br>Household heating: efficiency standards, phase out, scrapping<br>Vehicle standards, biofuel mandates |
| Public procurement | Public investment | Public procurement for power generators, vehicles, heating devices, to kick-start diffusion |
| Monetary | Base interest rates | Lower interest loan programs |

*NOTE: The exogenous/endogenous phase-out terminology refers to whether the technology trajectory is specified or not. In FTT, the user decides on whether to impose a phase-out, and the model determines the trajectory.

In FTT, economic incentive policies influence the behaviour of the choice model. They come in the form of taxes, subsidies or feed-in tariffs, that are used to influence the costs that agents attempt to minimise. For example, capital cost subsidies in power generation influence the levelised cost of generating electricity for a particular technology, which then raises its attractiveness in the discrete choice model that is part of the replicator equation.

Policies in FTT can also be of regulatory form, in which case they restrict what the choice spectrum is for of the investor or consumer. For example, if vehicles of the current petrol engine generation are phased out, they cannot be chosen by agents, and will undergo an exponential decline as a result at a rate that is function of their survival. Vehicles can furthermore be scrapped. New types of vehicles can also be introduced in the market, through a purchase program, either funded or enforced by the public authorities, to kick-start a new technology market (e.g. regulating taxi companies with respect to their vehicle efficiency). Finally, the content of liquid fuels can be changed by regulation through biofuel mandates.

### *3.1. Coupling of the models to one-another*

Coupling between the macroeconomy and technology systems is crucial, and this is done dynamically simply by integrating the E3ME and FTT models into the same computer code. Many feedback mechanisms are allowed. For instance, in power generation, feedbacks include (1) electricity prices/demand, (2) investment, (3) fuel use, (4) government income and expenditure on taxes and subsidies. The models are solved together iteratively. This includes the fossil fuel depletion model and FTT models for power, road transport and household heating.[6] The coupling between the GENIE1 carbon cycle/climate is soft-linked to E3ME-FTT, in order to lower computational demands, as discussed in the SI.[7]

## 4. Discussion and policy implications

We apply the model here as an example by exploring the economic impacts of an elaborate bundle of policies aimed at generating a low-carbon transition that achieves the goals set by the Paris Agreement. In the next section, we list the details of the chosen policies, and following that, we explore the technological and economic implications.

[6] FTT models for industry (iron & steel, other metals, chemicals, etc) and for agriculture/land-use are under active development. FTT:Agriculture will establish the link between the climate and the economy.

[7] This will change as we attempt to study problems of deforestation, in which the economy directly influences the climate.

### 4.1. Scenario for >70% chance 2 ℃

We provide here an example of a basket of policies that enables, in the E3ME-FTT model, to achieve emissions reductions consistent with greater than 66% probability of not exceeding 2°C of global warming. We note that bioenergy with carbon capture and storage (BECCS) is not a dominant feature of our scenarios, even if the consequence is higher system cost overall. We stress here that all of the policies included play a role in the broader emissions trajectory. We showed elsewhere [4] that policies interact and that the sum of their impacts can be greater than the sum of the impacts of policies applied individually. We do not claim, however, that this is the only basket of policies that can achieve the goals of the Paris Agreement. We note that these policies are added to the baseline case, and that policies in the baseline scenario are mostly defined implicitly through the data that was used. This is the case since trajectories of diffusion of innovations, as observed in our historical data, take place partly due to existing policies (e.g. existing transport policies or incentives for households), on which we have no information, and thus are represented implicitly in the model.

**Electricity sector (FTT:Power model)**

- **Feed-in-Tariffs -** 100% of the difference between the levelised cost for renewables and the spot price, plus a 10-20% additional incentive to promote renewable uptake (wind and solar only).
- **Direct subsidies** – up to 60% of the investment cost . Phased out by 2050
- **Carbon price** in all regions increasing gradually to 500$/tCO2 in 2050 (2008 dollars)
- **Regulations** are used to phase out or cap coal in some regions
- **Kick-start** for bioenergy with Carbon Capture and Storage (USA, China, India).

**Road transport sector (FTT:Transport)**

- **Standards** – more efficient internal combustion engine technologies are introduced as standard in 2017.
- **Regulations** are used to phase out older less efficient combustion engines.
- **Taxes on registration based on rated emissions**, of 100$/(gCO2/km) for every gCO2/km more than the lowest emissions category
- **Taxes on fuel**, increasing up to 0.50$/litre of fuel, in 2012 USD
- **Public procurement** – Electric vehicles introduced in the market in 2020 in all consumer categories
- **Biofuel blend mandate** that increases over time, starting at current levels, reaching 97% in 2050.

**Household heating:**

- **Fuel tax** of 50€/t$CO_2$ in 2020, increasing to 150€/t$CO_2$ in 2050
- **Subsidies** of 50% of the capital cost for renewable heating systems
- **Kick-start** for low-carbon technologies with no presence in various regions

**Other sectors**

- **Regulations** – Coal phased out in China in non-power applications of heavy industry, replaced by electricity.
- **A biofuel blend** is assumed to increase by 10% per year in aviation
- **Regulations** – Household use of fossil fuels for heating regulated to decrease by 3% per year worldwide.

### *4.2. Technology diffusion for reaching 2 °C*

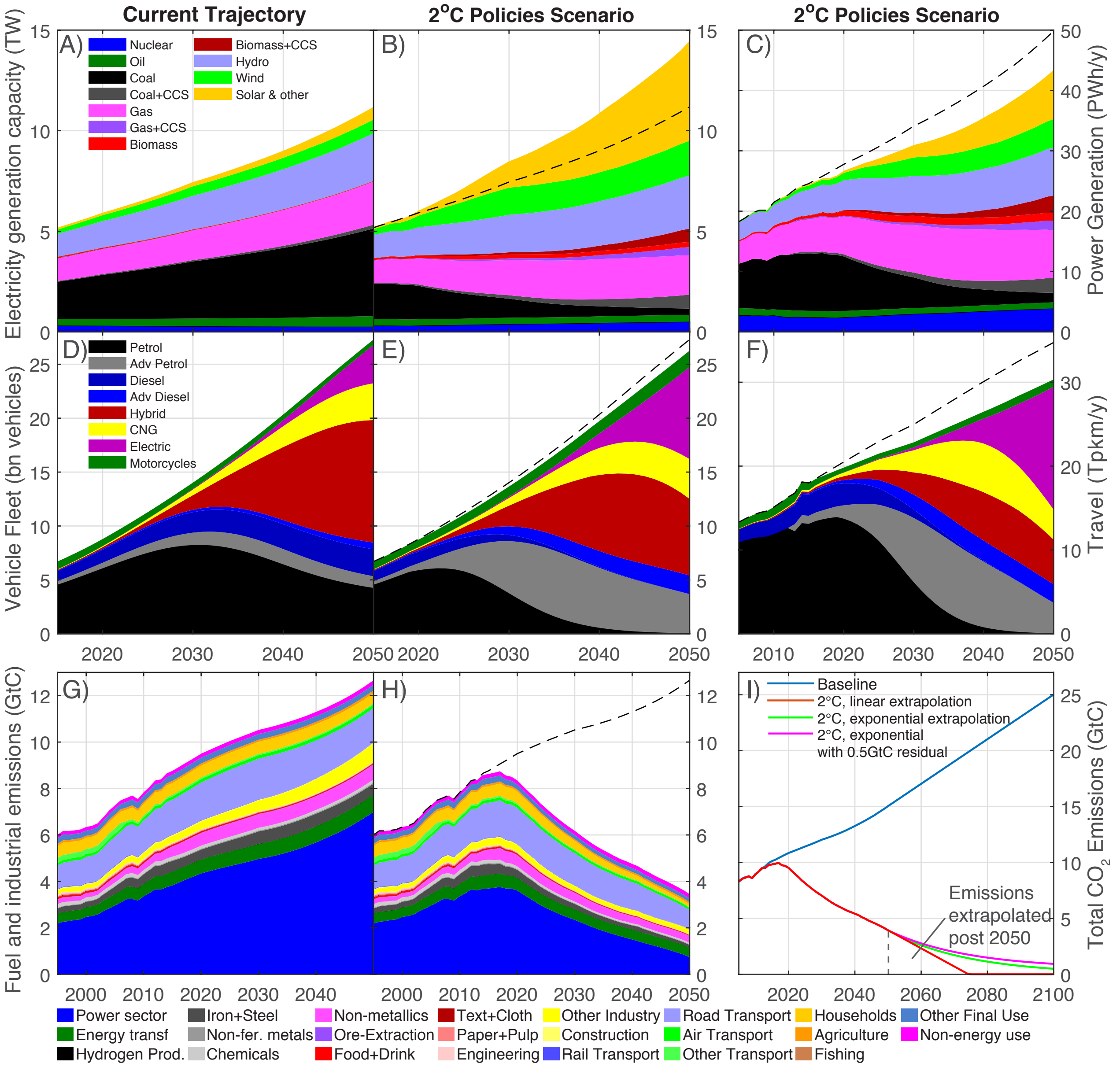

Figure 5: Technology diffusion in FTT:Power and FTT:Transport for a 2°C scenario in comparison to a current policies baseline, in terms of capacity (in GW). Panels A-C shows power sector capacity by type of technology. Panels C-F show the technology composition of the vehicle fleet. G-H show total fuel and industrial emissions by sector. Panel I) shows emissions trends extrapolated to 2100 for use in GENIE1. Column A-D-G shows the FTT baseline scenario, column B-E-H shows the 2°C scenario. Panels G and F show the 2°C scenario in units of service generation (in $10^{15}$ Watt-hours and $10^{12}$ person-kilometres). Freight transport, included road transport as a fuel user, is not shown here. Power and transport classifications were aggregated for clarity.

Figure 5 shows the diffusion of technology as a result of the policies listed above, in terms of capacity, in TW for the power sector and in vehicles for road transport. In the first column, the 'current policies' baseline is shown, with a development of the sectors that involves a slow diffusion of low-carbon technologies. In particular in the power sector, the current technology composition is mostly maintained, while for transport, higher efficiency vehicles (hybrids, natural gas) gradually replace lower ones.

The second column shows the impact, in power generation and transport, of policies given above. A faster diffusion of technologies is observed, including renewables in power generation and electric vehicles in transport. For transport, waves of ever higher efficiency vehicles arise one after the other before electric vehicles begin their mass diffusion. These changes lead to substantial changes in fuel use and emissions, since these sectors account together for over 60% of $CO_2$ emissions. These changes have economic impacts shown below. The third column shows energy service generation in that scenario (in GWh/y and Tpkm/y).

The combination of all sectoral contributions leads to substantial emissions reductions, sufficient to reach a probability greater than 70% of not exceeding 2°C of global warming.[8] This is shown in the lower row of panels in Figure 5, which gives global $CO_2$ emissions by fuel user. This emissions trajectory was used with the GENIE1 model to show the likelihood of meeting the target (see Figure 4). We consider it consistent with the PA.

Panel l) of Figure 5 shows how emissions were linearly extrapolated beyond 2050. This is reasonable because of the combination of negative emissions from BECCS and positive emissions lead to a linear trend. However, we assume that once emissions reach zero, the carbon price should decline substantially, and the business case for operating expensive negative emissions could become uncompetitive, and thus negative emissions stop soon after 2050. To test the reliability of this extrapolation, we provide other types of extrapolations: exponential decay to zero emissions, and exponential decay to residual emissions of 0.5 GtC, reflecting that some emissions sources could be challenging to remove. This reduces the probability of reaching the 2°C from 75% (linear) to 70% (exponential) and 68% (0.5 GtC residual), all of which achieve the goal of the PA. Thus, we do not expect that any other types of extrapolations lead to missing the target.

### *4.3. The economics of a 2 °C scenario in a simulation model*

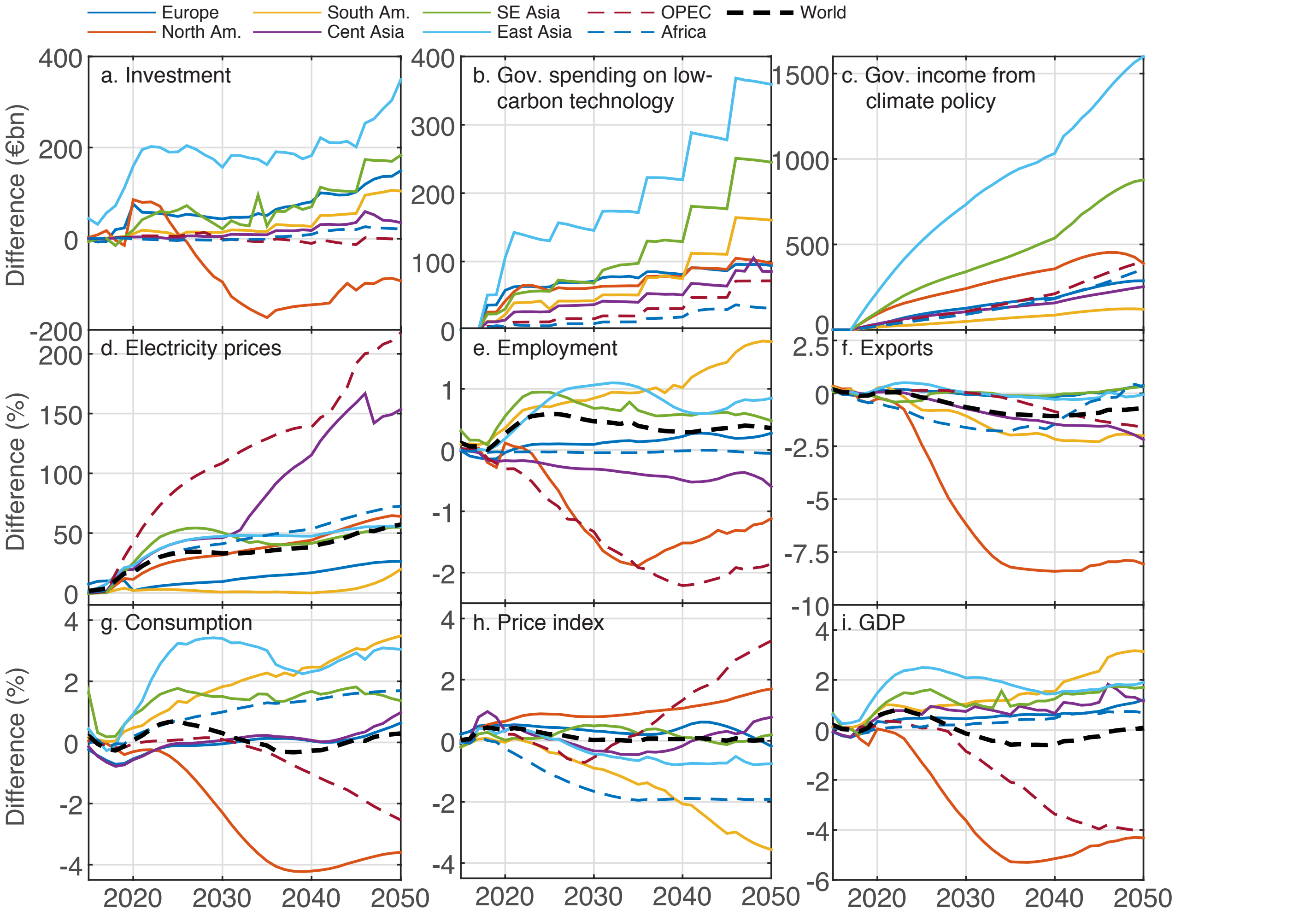

Figure 6: Economics of a low-carbon transition, viewed under a number of economic indicators, aggregated from 59 to 8 regions for clarity. In each panel, a difference to the baseline scenario is taken, and expressed in monetary values (panels a-c) or in percentage change (panels d-i).

Climate policy, leads to drastic changes in the use of fossil fuels. It also demands substantial investment into clean technology, and potentially large flows of money through public authorities to create the appropriate incentives for this investment to take place. This has important impacts to the economy, shown in Figure 6.

[8] 75% chance of not exceeding 2.0°C, 80% chance of not exceeding 2.04°C, with median 1.7°C.

Feedbacks to the economy operate in four ways. First, the costs of the transition are borne by consumers through higher energy service-related prices, and in particular, electricity prices. Higher production costs lead firms to increase their sale prices, which overall results in lower real disposable income for households. In general, this slows down regional economies. Secondly, investment in low-carbon technology, equipment and infrastructure, originating from increased total leverage (public and private debt) increases employment, thereby increasing household income. Thirdly, government spending typically stimulates regional the economies, while carbon taxing or pricing increases regional prices overall. However, income from taxes is typically larger than expenditures in each region and is eventually spent, for example for reducing income taxes, which can increase household income and consumption. Fourthly, declines in fossil fuel use reduces exports and income in fossil fuel-producing regions, decreasing their income, while phasing out imports of fossil fuels for non-producer regions improves their trade balance, raising their income. The timing of these effects is, however, not simultaneous.

Panel (*a*) shows changes in investment in comparison to the current policies baseline scenario, in absolute constant Euros (ref year = 2000). Investment is higher in a low-carbon scenario in countries where energy demand is growing, but not necessarily in regions where energy demand is stable (e.g. see data from [85]). In countries with growing energy demand, public spending is substantial (*b*), although more than covered by income from carbon taxing and pricing (*c*). The cost of the transition is partly covered by electricity consumers through higher electricity prices, that make up for higher costs of generation (d), which acts as a drag to the economy. However, investment and building activity generates employment (d), from which households enjoy increased disposable income (f). As a result, consumption is enhanced (g), despite increases in the price index (h), and GDP can be overall enhanced, depending on the balance of all factors.

Meanwhile, in countries where energy demand is stable, changes are comparatively small. This excludes fossil fuel producers (e.g. USA), for which the loss due to declining fossil fuel exports is larger than the income generated by low-carbon investment.

Differences between countries observed in Figure 6 can be explained as follows. Fast growing economies (e.g. China, India, Africa) receive intensified investment, as a fast growing trajectory of high carbon assets becomes re-directed into fast growing capital-intensive renewables, leading to employment, income and GDP increases. However, they suffer from increased energy prices. Meanwhile, fossil fuel exporters (e.g. USA, OPEC, Russia) suffer substantial stranded fossil fuel assets and declining investment. Fossil fuel importers (e.g. EU, China, India, Japan) benefit moderately from reduced fuel expenditures.

### 4.4. Comparison to other energy models

The E3ME-FTT-GENIE1 IAM differs quite significantly from most existing IAM models, for reasons given in this section. For reference to other models, we cite the 27th Energy Modelling Forum (EMF27) [86], the 5th Assessment Report of the IPCC (Working Group III, [19]), and earlier FTT model studies [4, 59]. Most differences to other models can ultimately be brought down to the simulation nature E3ME-FTT, as most other models use system-wide cost/utility-optimisation algorithms (the social planner assumption).

In terms of technology diffusion and composition, in particular for renewables, EMF27 shows large variations across models, which are ascribed to a number of factors [86, 87]: technology costs, technology availability, resource potentials, learning and power system integration. Costs and resource potentials are data source-dependent, and thus variations are to be expected. Some models have restrictions on availability of technology (e.g. solar PV on rooftops only) constraining the space of solutions. Only a subset of models include endogenous learning curves. Power system integration constraints are often included as fixed limits on technology shares (e.g. max 30% solar and wind), or included as additional integration costs.

Here, as a diffusion model, FTT functions quite differently. The diffusion pace is highly influenced by its own history, and therefore the diffusion process gains substantial momentum as it grows, a model property that we use to project current technological trajectories based on recent diffusion data. This explains the diffusion of hybrid and electric vehicles in the baseline scenario, without changes in policy, absent in other models, which is mostly driven by the fact that diffusion has been taking place in recent history, and the model only projects its continuation, assuming implicitly that current policies are maintained.[9] In addition, FTT features learning curves and fully endogenous power system stability constraints. Power system integration constraints, described elsewhere [59], change according to the system's composition. This ties, for instance, the diffusion of renewables with that of flexible systems in a sort of mutualism.

FTT results presented here differ from earlier reported results from the same model [4, 59]. Several improvements to the model have been made since these were published: higher regional resolution, higher policy resolution and improved cost and natural resources data. In particular, care was taken in recent scenarios to constrain the growth of the use of bioenergy and hydroelectricity to maintain these within more realistic bounds given existing debates [88], using regulatory policy. Furthermore, a much lower reliance on the carbon price is used, in line with existing proposed policy packages (notably in the EU). Data on costs were updated from 2008 to 2014, during which period solar PV and wind have seen spectacular progress and cost reductions, changing model outcomes substantially. We note that our model could not fully foresee this with 2008 data, underlying the inherent challenges of projecting technology markets. Model regional resolution also increased threefold.

### *4.5. Sources of variations in macroeconomic impacts of climate policies*

The differences in economic results from EMF27 and IPCC AR5 are more important than for technology systems, since they are often in contradiction. In general equilibrium models, investment is constrained by the amount of finance available, itself tied to the saving propensity parameters assumed in models, resulting in a fixed share of GDP available for investment. This equilibrium property implies that when higher than baseline investment is required for decarbonisation, the same amount of investment is cancelled elsewhere in the economy (crowding out), which by construction, always leads to GDP losses [17]. It implies, in some sense, that GDP can only be highest in the high carbon baseline, and economic impacts are thus expressed strictly as 'mitigation costs', excluding the possibility of negative costs [19, 86]. Meanwhile, in partial equilibrium models, system costs are obtained using the area under marginal abatement cost curves, themselves assuming that mitigation has positive costs. In all models, reported costs are roughly equal to the amount of investment required.

In E3ME-FTT, the baseline scenario is not special in any particular way, except in the sense that no additional policies are implemented, in comparison to today's world. It is not a scenario with necessarily lower energy system costs or higher GDP, or even lower investment. Economic impacts in E3ME relate mostly to financial and trade effects: changes in energy or other prices, employment, investment and in the trade balance. For instance, finance costs for investment in renewables are passed-on to consumers through electricity prices. Investment, however, generates employment. As opposed to general equilibrium models, E3ME does not assume money neutrality, but instead, models money creation by banks. Therefore, investment in one sector does not cancel out investment elsewhere, but instead, leads to higher leverage (total private debt). Higher than baseline investment generally leads to increases in price levels across sectors, in response to requirements to service debts incurred. Higher rates of inflation lead to lower real incomes and spending (depending on wage reactions) which leads to lower GDP but the change does not usually offset the positive effects of the investment stimulus. Including a financial sector to general equilibrium models could provide similar properties (although not all), as has been done with GEM-E3-FIT [17].

The pace of transformation, as opposed to its overall ambition, determines the magnitude of most economic impacts of climate policy in E3ME, in contrast to other models[10] One of the key detrimental impacts of climate policy is through stranded fossil fuel assets, in which fossil-related sectors shut down output, and substantial employment is lost. Thus, while importer countries see trade balance benefits, exporters suffer substantial GDP losses when the demand for fossil fuels declines. As finance is not a fixed share of GDP in E3ME, a lack of demand for investment in energy exporters is not compensated by shifts in prices and higher demand elsewhere, so there can be a large fall in overall production levels.

---

[9] Note that inertia in FTT is not only related to vintage effects (or turnover), but also, to assuming that technology availability and visibility increases with the state of diffusion.

[10] In standard equilibrium models, negative GDP impacts arise proportionally to total cumulated investment in low-carbon technology, while in E3ME-FTT, macroeconomic impacts arise with the pace of technological change, and the rate at which the economy can absorb these changes.

### 4.6. Deconstructing a basket of policies for 2 ℃

Since E3ME-FTT features a broad palette of possible policies for reducing emissions (section 4.1), substantial uncertainty can be associated with the particular composition of any policy basket. In order to estimate this uncertainty, it is necessary to run the model under many alternate baskets with small variations. Note that many baskets can reach the same emissions target, and that the full policy space has not yet been fully explored.

Here, we have produced a set of over 50 simulations to do this, shown in the SI. We show how global warming changes (expressed using 80% probability), when removing groups of policies in groups of countries from the 2ºC basket. We find that carbon pricing is the most important policy instrument, but that no country pulling out of climate policy on its own can increase warming to more than 2.7ºC. This, however, represents only the case for one policy/region, and since the model is non-linear, removing more is not linearly related to this. This is due to the fact that countries and policies interact, as shown in earlier work [4]. Note also that while the carbon tax appears to incentivise most of the decarbonisation, when assessed in terms of peak warming, this is a reflection of the dominance of the power sector for emissions; however, technology compositions in other sectors would not change substantially without other policies.

### 4.7. Sensitivity analyses: technology uncertainty

E3ME-FTT is a relatively stable model, by which we mean that it is robust against changes of input data. The model does not generate step changes or flips. Instabilities can be the result of faulty regressions, which we identify and remove. Being a path-dependent dynamical model, E3ME-FTT displays strong time autocorrelation, which means that every time step is naturally strongly related to its preceding time step.

The model features intrinsic output uncertainty through uncertainty over its parameter values. Since it is a dynamical model, uncertainty accumulates over simulation time span, which means that long-term outcomes can change substantially for small changes in input parameters (see e.g. [33]). However, this doesn't mean that the model is unstable over changes of parameters. In FTT, for large changes in technology uptake rates to take place (uptake is always continuous and changes smoothly), changes in mean perceived costs must substantially exceed the width of its associated distribution (see Figure 2). If not, changes are imperceptible. In the SI, we provide a sensitivity analysis over eight parameters of FTT:Power and FTT:Transport, by amounts that roughly correspond to the widths of the cost distributions. These parameters are those, other than policy, that generate the largest changes in emissions.[11] We find that for cost changes of 20%, technology shares change, by 2050, by at most 20%. We furthermore show the propagation of these variations into E3ME, expressing related changes to GDP and employment. The propagation of errors from E3ME-FTT to GENIE 1 can be estimated as the spread of possible outcomes in E3ME convoluted with the spread of outcomes of GENIE1 shown in Figure 4.

### 4.8. Summary for policy-making

We summarise here what has been learned for better informing policy-making. The model presented here is a descriptive model, as opposed to more common normative ones. Its usefulness lies in the detail of technology diffusion modelling combined with non-equilibrium economic modelling. For policy-making, what is key is the high policy space that this model offers.

We observe, for instance, that in such a model structure, most policies cannot be expressed in terms of a carbon price-equivalent, because policies interact with one another. For example, using a public procurement policy in transport, to kick-start the EV market for, enables taxes and subsidies to have a stronger effectiveness. Another example is where regulatory policy on what power generation technologies can be built, changes the effectiveness of the carbon price.

[11] Other parameters that could drive uncertainty, in the case of climate policy analysis, are energy price elasticities and technological progress indicators. These do not influence the technology composition, and thus influence emissions less, and are not analysed here.

The large policy space of this model, and interactions between policies, imply that identifying optimal strategies is not possible, as too many policy frameworks can lead to the same outcome. For example, one could take a more regulatory approach, or one could take a more market-based approach, and reach similar outcomes for emissions reductions. This is why this model is particularly suitable for impact assessment of detailed policy packages, but less so for agenda setting. We have shown here how a particular policy package in the model can achieve emissions reductions consistent with the Paris Agreement. However, other policy frameworks could also be identified that reach similar outcomes.

## 5. Conclusion: blueprint for a new role for integrated assessment models post-Paris

Designing a policy strategy to implement the Paris Agreement is a complex process that will involve trial and error. Time, however, is limited, and policy-makers must use all available information to ensure success. The policy cycle requires a detailed assessment of every component of a broader policy strategy in order to gain sufficient public and political support for it to be turned into law.

We presented here an integrated assessment modelling approach which is in many ways a first of a kind. It involves an integrated model simulation of the economy, technology and climate system with the highest available definition of policy instruments. It can be used to analyse in detail the likely impacts of complex baskets of low-carbon policies, and determine their ability to achieve policy objectives such as climate targets.

We stress that while modelling the future is highly uncertain, it is nevertheless the only method available to quantitatively inform policy-making. While the meaning of results can only be understood in the context of their respective uncertainty, we believe that the use of simplified models with the aim of generating simpler storylines is insufficient and could be misleading, while instead, the use of complex methods can improve our understanding of reality. Similarly, policy-making cannot be reduced to simple pricing-only strategies. The details of policy instruments matter, and their analysis is context-dependent, a reality that must increasingly be taken into account in models. We argue that this model can be used as a blueprint for the design of better models that could be used to analyse the impacts of low-carbon policies around the world.

## Acknowledgements

All authors acknowledge C-EERNG and Cambridge Econometrics for general academic and technical support. JFM, HP, PS, JV, NRE and PH acknowledge funding from the UK's research councils: JFM acknowledges funding from the Engineering and Physical Sciences Research Council (EPSRC), fellowship no. EP/ K007254/1; JFM, PS and JV acknowledge funding from two Newton Fund grants, no EP/N002504/1 (EPSRC) and ES/N013174/1 (Economic and Social Research Council, ESRC). NRE and PH acknowledge funding from the Natural Environment Research Council (NERC) grant no NE/P015093/1. Additionally, PS acknowledges funding from Conicyt. JFM and HP acknowledge funding from The European Commission's Horizon 2020 Sim4Nexus grant, and from DG ENERGY, and AL acknowledges a postdoctoral fellowship from the University of Macau. JFM acknowledges H. de Coninck and M. Grubb for informative discussions.

# Environmental impact assessment for climate change policy with the simulation-based integrated assessment model E3ME-FTT-GENIE

## Supplementary information online

**Jean-Francois Mercure**[1,2,3], **Hector Pollitt**[2,3], **Neil R. Edwards**[4], **Philip B. Holden**[4], **Unnada Chewpreecha**[2], **Pablo Salas**[3], **Aileen Lam**[3,5], **Florian Knobloch**[1] and **Jorge E. Vinuales**[3]

[1]Radboud University, Netherlands, Department of Environmental Science, Radboud University, PO Box 9010, 6500 GL, Nijmegen, The Netherlands

[2]Cambridge Econometrics Ltd., Covent Garden, Cambridge CB1 2HT, UK

[3]Cambridge Centre for Environment, Energy and Natural Resource Governance (C-EENRG), University of Cambridge, 19 Silver Street, Cambridge CB3 1EP, UK

[4]Environment, Earth and Ecosystems, The Open University, Milton Keynes, UK

[5]Department of Economics, Faculty of Social Sciences, Humanities and Social Science Building, University of Macao, E21, Avenida da Universidade, Taipa, Macao, China

Email: J.Mercure@science.ru.nl

## Detailed model information and sensitivity analysis

### 1.1. Basic model information

E3ME and FTT are time step path-dependent simulations. FTT runs as a discretised non-linear differential equation (a finite differences equation), each of which step is calculated from the previous time step. In order to avoid chaotic dynamics, the time step in FTT must be much smaller than the pace of change modelled. We found with experience that using a quarterly time step is sufficiently short to avoid chaotic dynamics while maintaining the simulation time manageable, without substantial loss of accuracy.

E3ME functions slightly differently from FTT, as it uses yearly time steps, with an error-correction procedure [1, 2]. In other words, once a year has been calculated, a long-run equation estimates the next year's value for each econometric equation, and solves iteratively with small changes using a short-run version of the same equation (see the E3ME manual [3]).

E3ME and the various FTT modules are solved iteratively. FTT modules typically have shorter time steps than E3ME, and therefore run their differential equations for a number of steps before reporting back prices and investment values to E3ME, which supplies a demand value. Through E3ME's iterative process, all econometric equations are solved with all FTT models, and iterations stop once value changes are maintained within a certain bound requiring between 50-100 iterations. Thus, FTT and the E3ME econometric equations are truly dynamically linked.

The GENIE1 model is linked with E3ME-FTT using a soft-link, as it is currently outside of our computing capacity to simultaneously run GENIE1 dynamically with E3ME-FTT. However, since timescales in the climate are much longer than those in the economy, the feedbacks would not need to be fed yearly between models, and in fact, the use of a soft coupling is sufficiently accurate. Even in a case where we would include some limited climate feedbacks to the economy (a full model cycle; for instance, through climate impacts on agriculture, back to climate), we expect that a relatively low number of iterations between models would be sufficient to obtain convergence, thus still supporting the use of soft-coupling. However, in the study of issues with strong feedbacks, such as deforestation (e.g. in the Amazon), where the economy interacts directly with the climate through changes in the water cycle; in this case, E3ME-FTT would require to be run dynamically with GENIE1, something that requires substantial additional work.

E3ME-FTT-GENIE1 is not currently available publically to download; however, for the interest of readers, it is possible to use the model in collaboration with the authors. The reasons are as follows. Firstly, it takes substantial amounts of resources to train users, as well as substantial dedication from users (over half a year of experience is typically necessary to become a developer, and Cambridge Econometrics offers a one week course for light users). Secondly, the models are large and bulky to transfer to users; and GENIE1 runs on a university cluster. We recommend interested readers to contact us at hp@camecon.com (H. Pollitt), J.Mercure@science.ru.nl (J.-F. Mercure) and/or Phil Holden (philip.holden@open.ac.uk).

### 1.2. Sensitivity analysis and uncertainty propagation across E3ME-FTT

We provide here an analysis of the sensitivities that arise from varying FTT technology parameters, and how they propagate to E3ME and back to FTT across all feedbacks. We varied all parameters we considered key for varying the quantities of fossil fuels consumed and emissions generated, in both scenarios. Results are given below in Tables 1-2. We varied technology costs, consumer (transport) and corporate (power generation) discount rates, learning rates, non-pecuniary costs (transport) and the fuel efficiency of combustion vehicles (transport). Results are expressed as % changes of a particular quantity over its original value in both scenarios. We find that technology shares in 2050 vary by relatively small amounts when parameters change by values of the order of 20% (costs) or 5% (discount/learning rates). Values change less in the 2ºC scenario than in the baseline, a reflection of the relative stability that climate policy provides (e.g. see the shares of solar PV). This demonstrates the reliable stability of the E3ME-FTT model. Note that these changes arise only gradually between 2017 and 2050. E3ME-FTT being a non-linear model, it 'cumulates' differences over time, which can become large with a long simulation time span. This remains however within reasonable bounds, as the data shows in Tables 1-2.

We explored the impact on the macroeconomy of these variations on technology parameters. We use GDP and employment as indicators, which are cumulated between 2017 and 2050. Cumulated changes in unemployment can be interpreted as job creation or job loss. Changes remain within the order of 1% or below. Note that a global GDP change of 1% is a relatively large quantity of production, considering that power and road transport only account for a relatively small component of GDP. USA GDP is affected more than China or the EU primarily through losses of fossil fuel production. Combined uncertainties across scenarios are given using the root of the sum of the squares of the variations (assuming equal probabilities).

Other parameters that could drive uncertainty, in the case of climate policy analysis, are energy price elasticities and technological progress indicators. Uncertainty over these parameters could be significant for calculating GDP but less so for emissions. Note that part of the reductions in energy demand taking place with higher energy prices stem from endogenous investment in energy efficiency, which does not depend on energy price elasticities. A full sensitivity analysis for these parameters is a substantial undertaking, outside of the scope of the present work.[1] Uncertainty propagation between E3ME-FTT and GENIE1 on peak warming is straightforward to evaluate, given that for small variations in cumulative emissions, climate patterns do not change. Therefore, the spread of values given in Tables 1-2 can be translated to peak warming variations by convolving them to the spread of values given in Figure 4 of the main text.

### 1.3. Varying the basket of policies

We varied the basket of policies discussed in the main text, in order to examine the role of each item. This work only covers broad groups of policies and groups of countries; an in-depth analysis lies outside of the scope of this paper, as it can make a separate publication on its own. In this analysis, over 50 simulations were carried out, in which every policy instrument group is taken out in every regional group. Policy groups include the carbon tax/price (denoted 'Tax'), road transport ('RT'), publicly funded energy efficiency ('EE'), household heating ('Heat') and regulations ('Reg'). The impact on peak warming, expressed with an 80% probability, is shown in Figure 1.

[1] E3ME has tens of thousands of econometric regression parameters, due to its high level of disaggregation. We estimate that carrying out a full sensitivity analysis such as we present for GENIE1 would take several decades of CPU time, perhaps even up to 300 years. An efficient statistical method to carry out this task requires to be developed alongside a choice of subset of parameters to analyse.

| | Current Policies Parameter | Var % | REN % | PV % | EV % | ADV % | FF % | GDP GL% | EMP GL% | GDP US % | GDP CN % | GDP EU % |
|---|---|---|---|---|---|---|---|---|---|---|---|---|
| Power Generation | REN capital costs | +20 | -7.36 | -11.3 | 0 | 0 | 0 | 0.05 | 0.02 | 0.03 | 0.04 | 0.10 |
| | REN capital costs | -20 | 8.25 | 15.9 | 0 | 0 | 0 | -0.03 | -0.01 | -0.07 | -0.01 | -0.03 |
| | REN learning | +5 | 8.86 | 31.9 | 0 | 0 | 0 | 0.03 | 0.01 | 0.05 | 0.04 | 0.05 |
| | REN learning | -5 | -7.87 | -26.4 | 0 | 0 | 0 | 0.03 | 0.01 | -0.02 | 0.02 | 0.08 |
| | Discount rate | +5 | 3.96 | 51.1 | 0 | 0 | 0 | 0.36 | 0.09 | 1.08 | 0.45 | 0.06 |
| | Discount rate | -5 | 8.29 | -13.3 | 0 | 0 | 0 | -0.10 | -0.04 | -0.57 | -0.23 | 0.13 |
| Road Transport | Perceived costs | +20 | 0.60 | 3.08 | -5.51 | 0.61 | 1.21 | 0.09 | 0.01 | 0.30 | 0.05 | 0.02 |
| | Perceived costs | -20 | -0.53 | -2.38 | 15.0 | -3.33 | -0.48 | -0.08 | 0 | -0.26 | -0.03 | -0.03 |
| | Learning rates | +5 | 0.32 | 1.93 | 3.50 | 3.55 | -7.42 | 0.01 | 0.01 | 0.02 | 0.06 | 0.03 |
| | Learning rates | -5 | 0.03 | 0.07 | -10.1 | -2.99 | 9.18 | 0 | 0 | -0.01 | -0.03 | -0.01 |
| | Discount rate | +10 | -0.02 | -0.04 | 6.89 | -0.01 | -2.78 | -0.02 | 0 | -0.09 | -0.01 | 0.01 |
| | Discount rate | -10 | 0.43 | 2.21 | -8.94 | 0.20 | 3.31 | 0.04 | 0 | 0.11 | 0.02 | 0.03 |
| | EV costs | +20 | -0.48 | -2.77 | -9.48 | 1.74 | 0.92 | -0.01 | 0 | 0.05 | -0.01 | -0.08 |
| | EV costs | -20 | 0.05 | 0.28 | 8.44 | -1.56 | -0.80 | 0 | 0 | 0 | 0 | 0.01 |
| | ADV Fuel Eff | +20 | 0 | 0 | 0 | 0 | 0 | 0 | 0 | 0 | 0 | 0 |
| | ADV Fuel Eff | -20 | 0 | 0 | 0 | 0 | 0 | 0 | 0 | 0 | 0 | 0 |
| | Root mean square | | 18.66 | 70.11 | 25.67 | 6.21 | 12.69 | 0.40 | 0.10 | 1.29 | 0.51 | 0.23 |

Table 1: Results of a sensitivity analysis for key technology parameters in the Current Policies scenario, expressed as % change over the original value in the scenario. Abbreviations: REN = shares of renewables + nuclear, PV = shares of solar photovoltaic, EV = shares of electric vehicles, ADV = shares of new, higher efficiency combustion vehicles, FF = shares of conventional combustion vehicles, GDP GL = global GDP, EMP GL = global employment, GDP US, CN, EU = GDP of the USA, China and the sum of EU member states. Technology parameters are taken in 2050. GDP and employment values are cumulated between 2017 and 2050.

| | 2°C scenario Parameter | Var % | REN % | PV % | EV % | ADV % | FF % | GDP GL% | EMP GL% | GDP US % | GDP CN % | GDP EU % |
|---|---|---|---|---|---|---|---|---|---|---|---|---|
| Power Generation | REN capital costs | +20 | 0.11 | 4.02 | 0 | 0 | 0 | 0.03 | 0 | 0.09 | 0.03 | 0.01 |
| | REN capital costs | -20 | -0.19 | -5.29 | 0 | 0 | 0 | 0 | 0 | -0.01 | 0 | 0 |
| | REN learning | +5 | -0.20 | -2.55 | 0 | 0 | 0 | 0 | 0 | 0.01 | 0 | 0.02 |
| | REN learning | -5 | 0.23 | 1.74 | 0 | 0 | 0 | 0 | 0 | 0.06 | 0.03 | -0.04 |
| | Discount rate | +5 | -3.09 | 7.20 | 0 | 0 | 0 | 0.05 | 0.05 | 0.41 | 0.22 | 0 |
| | Discount rate | -5 | 3.62 | -8.51 | 0 | 0 | 0 | -0.02 | -0.02 | 0.15 | -0.05 | 0 |
| Road Transport | Perceived costs | +20 | -0.01 | -0.60 | -3.37 | 0.29 | 9.60 | -0.03 | -0.01 | 0.07 | 0.01 | -0.14 |
| | Perceived costs | -20 | -0.05 | 0.29 | 11.9 | -8.47 | -18.3 | 0.02 | 0 | 0 | 0.02 | 0.07 |
| | Learning rates | +5 | 0 | -0.17 | 0.88 | -0.08 | -7.93 | -0.03 | -0.01 | 0 | -0.01 | -0.08 |
| | Learning rates | -5 | -0.13 | -0.25 | 2.24 | -8.93 | 4.95 | 0.02 | 0 | 0.05 | -0.01 | 0.04 |
| | Discount rate | +10 | -0.18 | -0.23 | 9.80 | -12.8 | -5.78 | 0.02 | 0.01 | 0.04 | -0.01 | 0.04 |
| | Discount rate | -10 | 0.10 | 0.27 | -10.5 | 11.7 | 4.70 | -0.02 | -0.01 | 0.03 | 0.01 | -0.04 |
| | EV costs | +20 | 0.01 | 0.01 | -2.30 | 1.16 | 0.07 | 0 | 0 | -0.01 | 0 | 0 |
| | EV costs | -20 | -0.01 | -0.01 | 2.55 | -1.26 | 0.19 | 0 | 0 | 0.01 | 0 | 0 |
| | ADV Fuel Eff | +20 | -0.02 | 0.12 | -0.56 | 0.41 | -2.48 | -0.03 | 0 | -0.01 | -0.01 | -0.07 |
| | ADV Fuel Eff | -20 | 0.02 | -0.11 | 0.77 | -0.48 | 6.75 | 0.04 | 0 | 0.04 | 0.01 | 0.07 |
| | Root mean square | | 4.78 | 13.36 | 19.48 | 21.35 | 24.93 | 0.18 | 0.06 | 0.47 | 0.23 | 0.21 |
| | Scenarios combined | | 19.26 | 71.37 | 32.23 | 22.23 | 27.97 | 0.44 | 0.12 | 1.38 | 0.56 | 0.31 |

Table 2: Results of a sensitivity analysis for key technology parameters in the 2°C scenario, expressed as % change over the original value in the scenario. Same abbreviations.

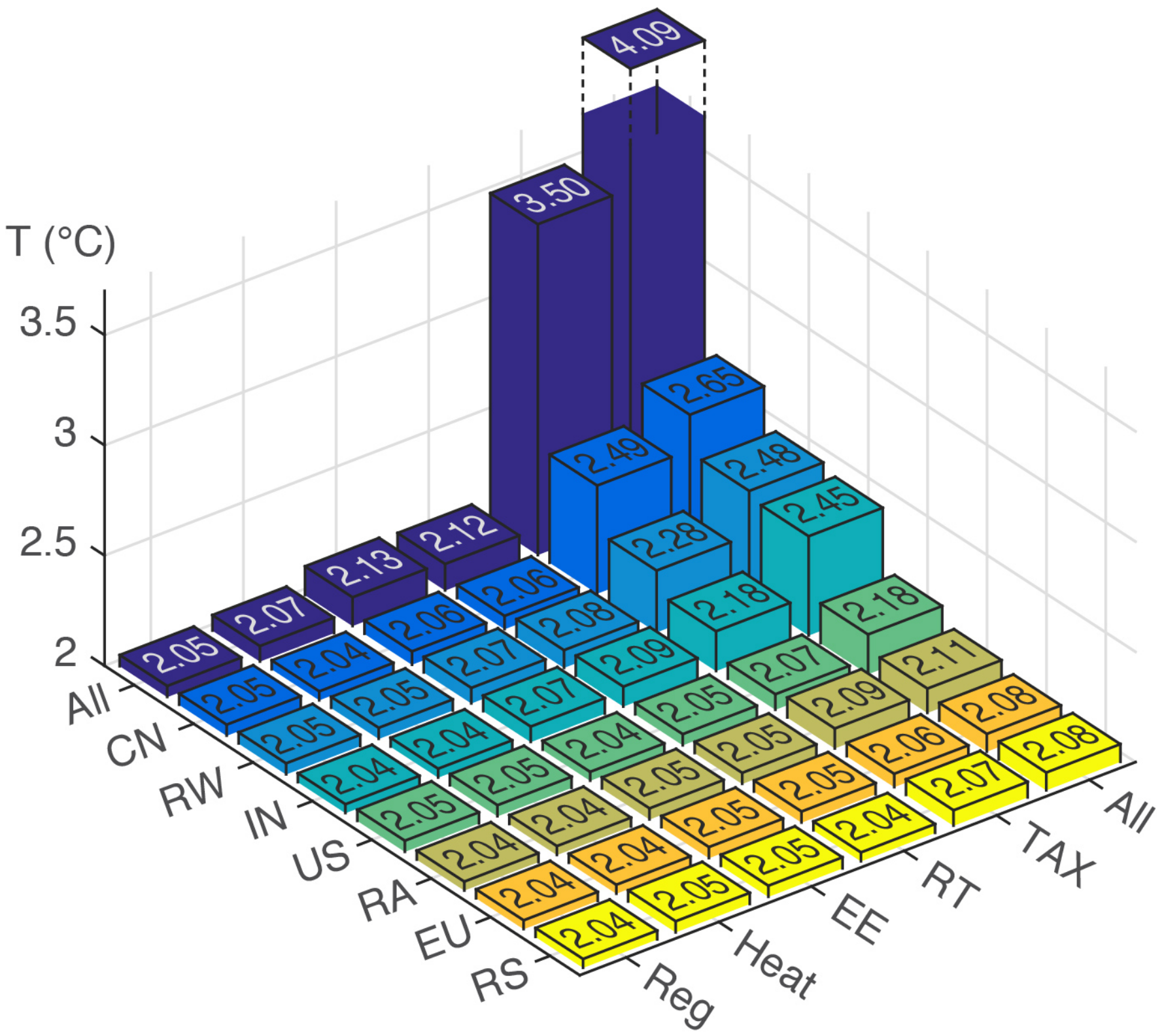

Figure 1: Impact of elements of the policy basket for achieving the 2ºC target. The vertical quantity refers to peak warming when calculated with an 80% probability (2.04ºC at 80% is equivalent to 2.00ºC at 73% according to GENIE1 simulation. Note that we used 0.0019ºC/tC to generate the variations in peak warming). CN = China, RW = rest of the developing World excl. regions specified, IN = India, US = United States, RA = rest of Annex 1 excl. regions specified, EU = European Union and RS = Russian Federation.

### *1.4. Econometrics in E3ME*

The econometric techniques used to specify the functional form of the equations are the concepts of cointegration and error-correction methodology, particularly as promoted by Engle and Granger [1] and Hendry et al [2]. In brief, the process involves two stages. The first stage is a levels relationship, whereby an attempt is made to identify the existence of a cointegrating relationship between the chosen variables, selected on the basis of economic theory and a priori reasoning, e.g. for employment demand the list of variables contains real output, real wage costs, hours-worked and measures of technological progress.

If a cointegrating relationship exists then the second stage regression is the error-correction representation. It involves a dynamic, first-difference, regression of all the variables from the first stage, along with lags of the dependent variable, lagged differences of the exogenous variables, and the error-correction term (the lagged residual from the first stage regression). Due to limitations of data size, however, only one lag of each variable is included in the second stage.

Stationarity tests on the residual from the levels equation are performed to check whether a cointegrating set is obtained. Due to the size of the model, the equations are estimated individually rather than through a cointegrating VAR. For both regressions, the estimation technique used is instrumental variables, principally because of the simultaneous nature of many of the relationships, e.g. wage, employment and price determination. The instruments used are the previous year's data, which is a standard approach in time-series econometrics.

We show in Figure 2 projections for the current trajectory and 2C scenarios in absolute values, i.e. not taking differences, using the same quantities in the same panels as Fig. 6 of the main text. The reader can see what baseline projections are for most relevant economic variables. Figure 3 shows projections for energy demand and supply for several energy carriers for both scenarios.

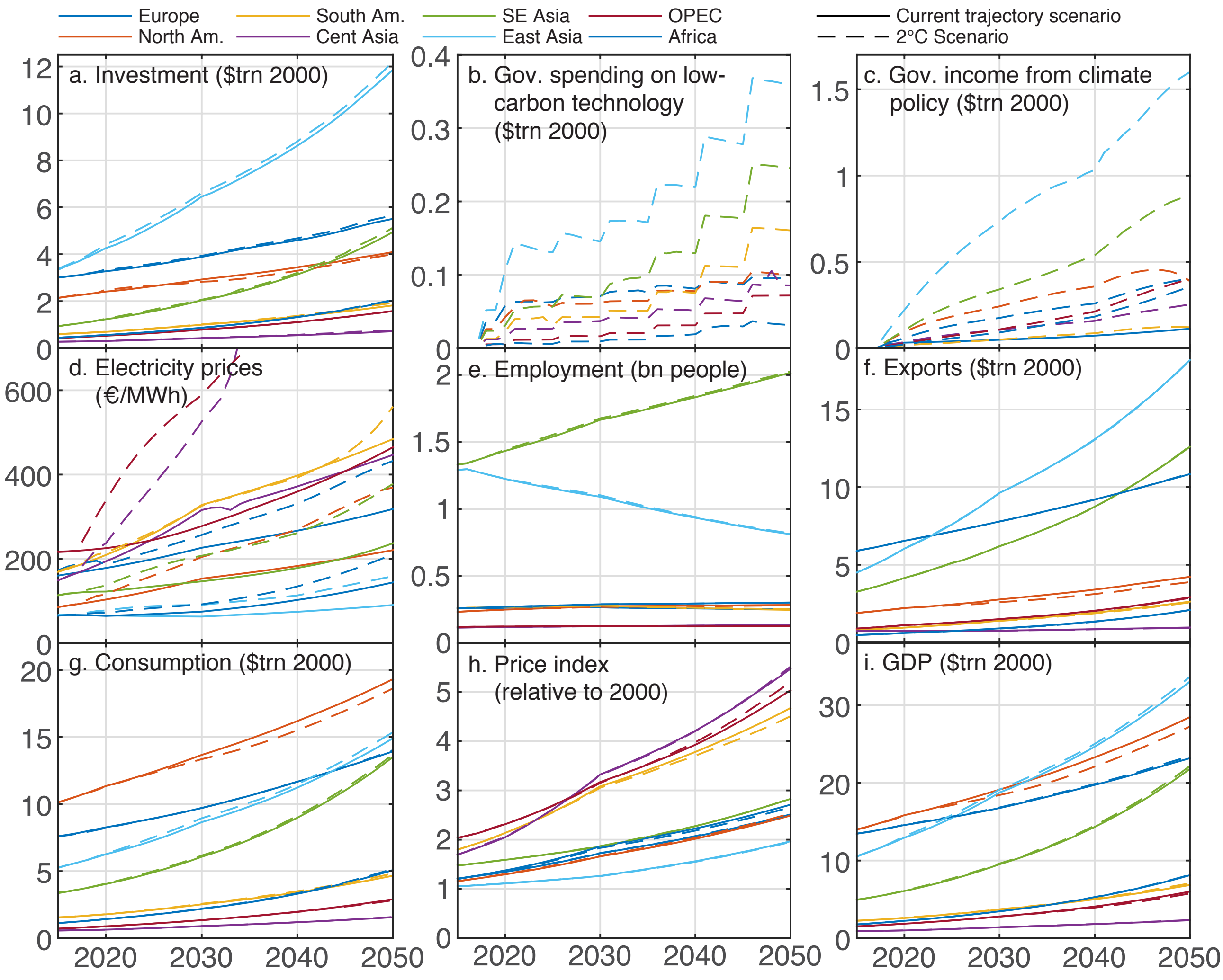

Figure 2: Projections for the current trajectory (solid lines) and the 2ºC scenario (dashed lines) when not taking differences (panels show the same quantities as fig. 6 of the main paper.

### *1.5. Calibration procedure*

The system of calibration in E3ME is quite different to that which is sometimes applied in CGE modelling. In E3ME the aim of the calibration is to allow the model to match a published baseline, while in CGE modelling it is used to determine model parameters. There is a crucial difference when comparing results from different scenarios: in E3ME the effects of the calibration should largely cancel out, while the calibrated parameter values used in CGE modelling can be a key determinant of results.

Model calibration in E3ME follows a specific procedure. First, the baseline forecast is stored on the model forecast databank. This process is carried out independently of the model itself. Second, the model is run in calibration mode, which means that the equations are solved, but the results are set to match the values on the databank. The differences between the model results pre- and post-calibration are stored on another model databank. In most cases the values are stored as scaling factors in the form of ratios, i.e. multiplicative differences. When the model is then solved as fully endogenous, the scaling factors are applied to the results of the equations, both in the baseline case and any scenarios. As the same scaling factors are applied in each case, when comparing the results between model runs, the influence of the scaling factors cancels out.

There are still, however, cases where the values in the baseline are important:

1- If a scenario has an absolute target, then the value of the baseline determines the level of ambition required to meet that target.
2- If a baseline is highly biased, then it could have some influence on model results (e.g. a scenario that reduces exports will have more impact if the baseline has lots of exports).

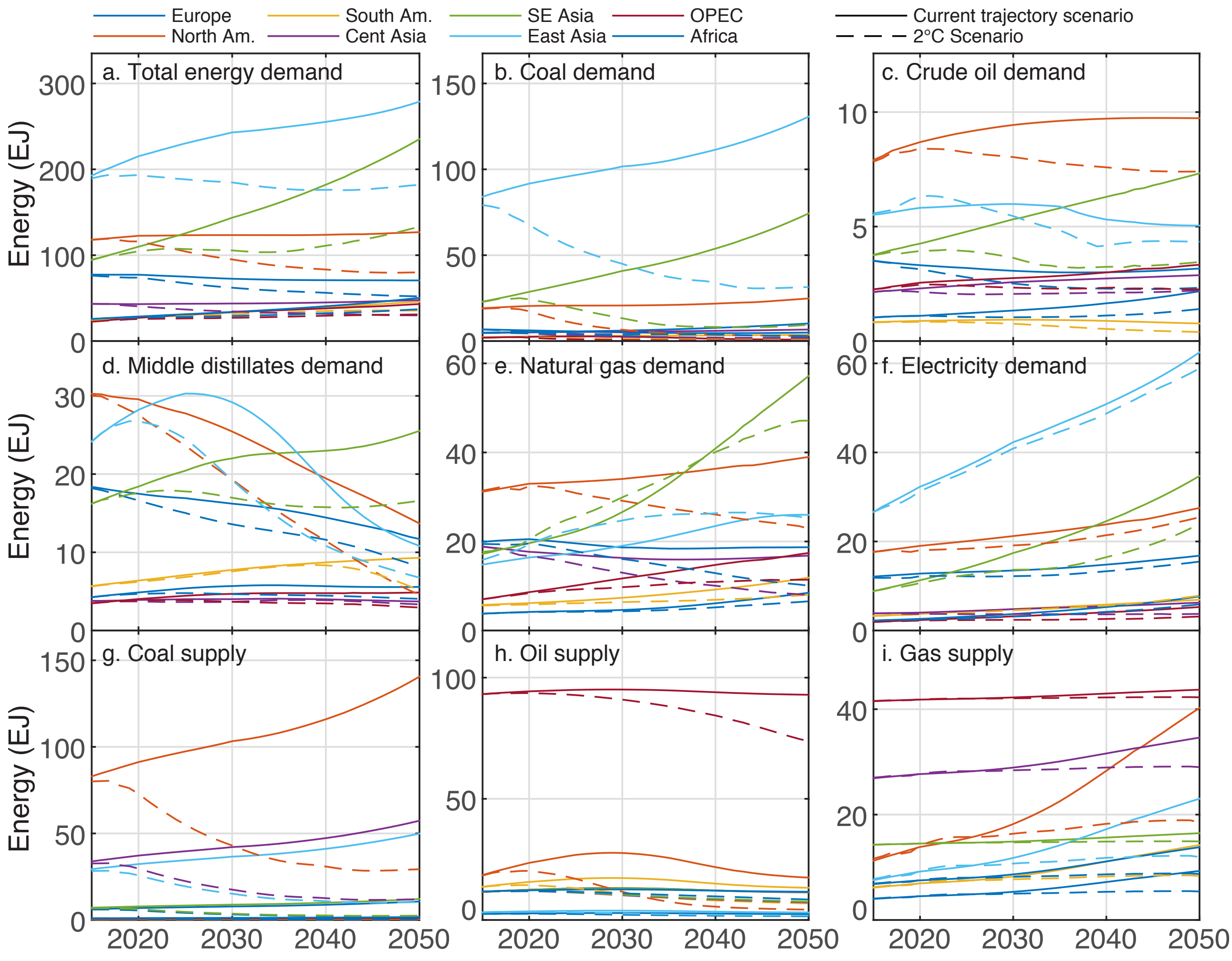

Figure 3: Energy demand (a-f) and production (g-i) by carrier in the current trajectory (solid lines) and 2ºC scenarios (dashed lines) in absolute values. Differences in totals consist of losses between primary and final consumption.

It is therefore important to form a baseline that does not introduce bias into the results. This baseline could be produced by the model itself but typically an externally-validated source is used. The most common source is the IEA's World Energy Outlook scenarios, which are based on OECD growth projections.

The exact values taken for calibration depend on the sources used. Key economic values include GDP, sectoral GVA and household expenditure. Other model variables are then adjusted to be consistent (e.g. sectoral gross output and investment follow similar growth paths to GVA). Other model variables are adjusted so that accounting balances are maintained. Employment projections are generally set to be consistent with the working age population, as provided by UN population forecasts. For model variables where there are no published projections to follow (notably trade and non-energy prices), historical trends are extrapolated and constrained to remain within reasonable boundaries.

Energy balances and prices are taken from IEA figures and further disaggregated (based on constrained extrapolations) to match the model classifications. Emissions may be either calibrated to projections, or solved endogenously based on energy consumption.

Some modelling exercises require sensitivity to test the importance of the baseline assumptions. The results from this analysis usually show that the exact baseline specification does not have a major influence on results (as % difference from baseline), as long as the values chosen remain within reasonable bounds.

Note that FTT outputs (e.g. technology shares within a sector) are not calibrated, but total demand values for energy services (e.g. electricity, transport) are calibrated.

### 1.6. Forecasting performance of E3ME

Tests on the forecasting performance of an earlier version of E3ME (then called E3MG) have been done and published in a chapter of the book edited by Barker & Crawford-Brown [4]. The test involved evaluating the econometric parameters as described above, although without the use of residuals, and then attempting to reproduce history between 1970 and 2007. Despite issues with data missing in various regions in various time spans (e.g. the Soviet Bloc), this was relatively successful. Some results are shown below in Figure 4, in which the model 'forecasts' are displayed against the actual data over history.

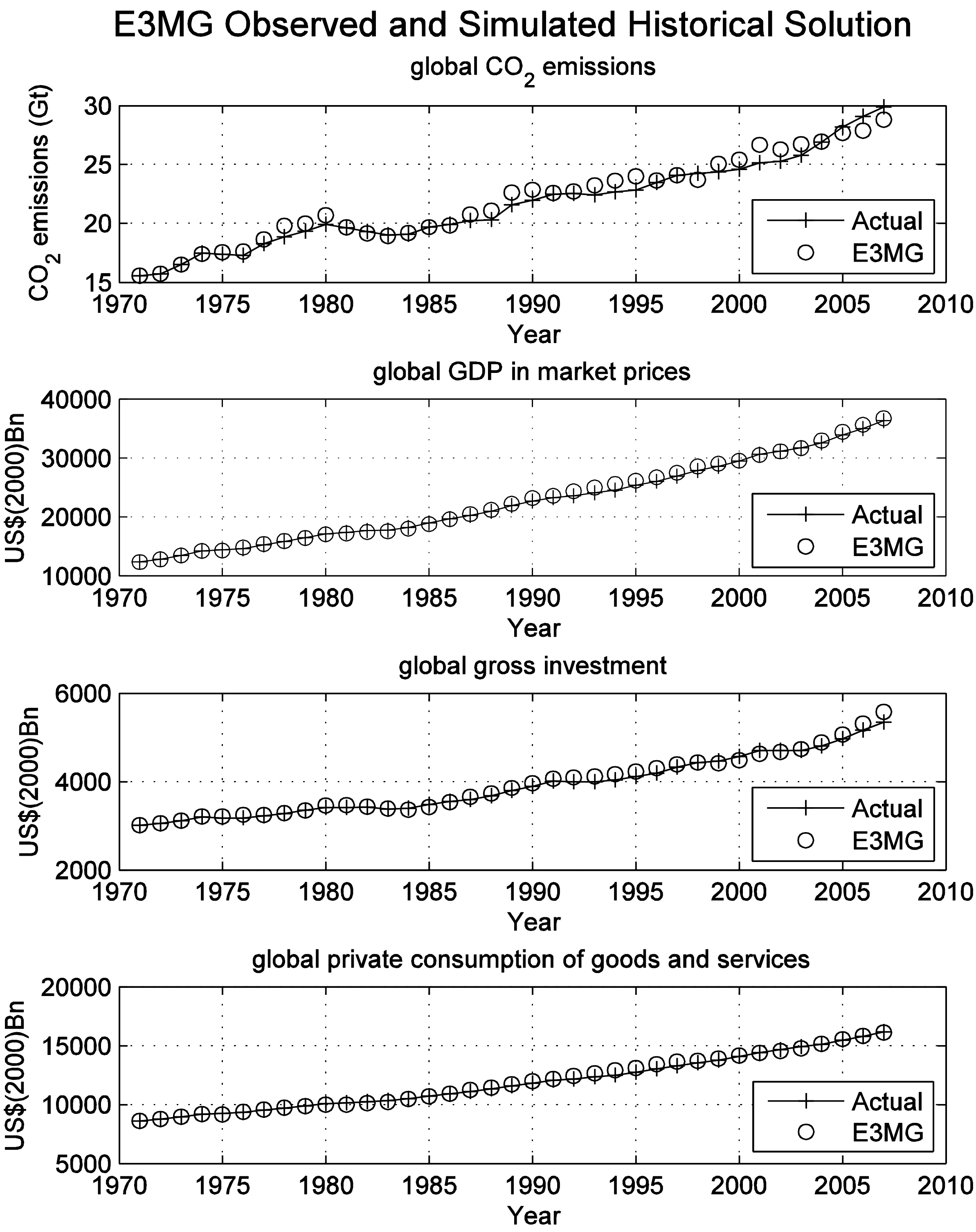

Figure 4: Computational experiment with the earlier version of E3ME (called E3MG at the time) in which the model is used to reproduce historical economic and emissions data without calibration, between 1971 until 2007. Reproduced from Barker & Crawford-Brown [4].

## Detailed cross-referenced model equations for E3ME-FTT

### 1.7. E3ME Model: key economic identities

In the rest of this document, we provide a complete list of model equations for E3ME-FTT. Brackets indicate reference links to: **[1]** econometric equations, (i1) identities, {x1} exogenous values and ⟨1⟩ conversion matrices. Further links connect E3ME to FTT variables: [f1] refer to dynamical differences equations, (fi1) to FTT identities and {fx1} to FTT exogenous (usually policy) quantities. This enables the reader to track model information flow by automatically searching the document.

**GDP identity**

While GDP provides a measure of net production at the whole-economy level, at the sectoral level we have (gross) output and gross value added. Output is equivalent to turnover in that it includes intermediate inputs to production, while value added does not include purchases from other sectors. RGDP, however, is not used in the model; it is only observed by the user.

RGDP = RSC + RSK + RSG + RSX - RSM

| | |
|---|---|
| RGDP(reg) | GDP, 2005 prices |
| RSC(reg) | total consumer expenditure, 2005 prices *(i59)* |
| RSK(reg) | total investment (GFCF), 2005 prices *(i60)* |
| RSG(reg) | total final government expenditure, 2005 prices *(i61)* |
| RSX(reg) | total exports, 2005 prices *(i62)* |
| RSM(reg) | total imports, 2005 prices *(i63)* |

**Calculation of output**

The measure of output is determined from the demand side, in a similar way to GDP but also including the intermediate demands, as shown below. Each variable in the box is defined by both region (i.e. state) and sector.

(i21) QR = QRY + QRC + QRK + QRG + QRX - QRM

| | |
|---|---|
| QR(reg,ind) | GDP, 2005 prices |
| QRY(reg,ind) | vector of output (by product), 2005 prices *(i75)* |
| QRC(reg,ind) | vector of final consumer output goods, 2005 prices *(i76)* |
| QRK(reg,ind) | vector of final investment goods, 2005 prices *(i78)* |
| QRG(reg,ind) | vector of final government goods, 2005 prices *(i77)* |
| QRX(reg,ind) | vector of final exported goods, 2005 prices *(i28)* |
| QRM(reg,ind) | vector of final imported goods, 2005 prices *(i29)* |

**Balancing supply and demand**

A fundamental part of the national accounting structure is that supply and demand must match. In the demand-driven structure of E3ME this is imposed by ensuring that production matches the level of the goods demanded (if there are supply constraints that prevent this from happening then demand must be adjusted separately).

(i22) YR = QR

| | |
|---|---|
| YR(reg,ind) | vector of output (by industry), 2005 prices |
| QR(reg,ind) | vector of output (by product), 2005 prices *(i21)* |

**Calculating value added**

Value added is defined as the difference between output and material input costs. Value added itself is the sum of wages, company profits and production taxes.

YRF = YR(ind) * (1 – YRUM(ind) – YRUT(ind))

| | |
|---|---|
| YRF(reg,ind) | vector of value added, 2005 prices |
| YR(reg,ind) | vector of output (by industry), 2005 prices *(i22)* |
| YRUM(reg,ind) | vector of unit intermediate costs by industry, 2005 prices *(i34)* |
| YRUT(reg,ind) | vector of unit tax costs by industry, 2005 prices *(i37)* |

**Consumer prices**

GDP and value added are among the most important model results but there are other identity relationships that play an important role in determining these results. The key ones are presented in the following paragraphs, starting with the measures of consumer prices and inflation.
Consumer prices are determined by converting industry prices to the relevant consumer products. For example, the prices of cars are determined by the output prices of the car industry, plus the contribution from transport and retail costs, plus the taxes on purchases of new cars.

(i72) PCR = QCC * PQRD * (1+CRTR)

| | |
|---|---|
| PCR(reg,con) | vector of consumer prices, by product, 2005 prices |
| QCC(reg,ind,con) | matrix that converts industry production to consumer products ⟨3⟩ |
| PQRD(reg,con) | vector of prices of industry sales to the domestic market, 2005 prices *(i39)* |
| CRTR(reg,con) | vector of indirect tax rates on consumer products {x25} |

### The consumer price index

The aggregate consumer price index is obtained by taking the sum across all consumer products. Inflation is the annual change in the consumer price index.

(i80) PRSC = sum (PCR * CR) / RSC

| | |
|---|---|
| PRSC(reg) | aggregate consumer price index, 2005 = 1.0 |
| PCR(reg,con) | vector of consumer products' prices, 2005 = 1.0 *(i72)* |
| CR(reg,con) | vector of expenditure on consumer products, 2005 prices *[1]* |
| RSC(reg) | sum of expenditure on consumer products, 2005 prices *(i59)* |

### Calculating nominal incomes

Real incomes are the main driver of consumption, which is the largest component of GDP. The level of real incomes is therefore a key model result. The variable is determined by summing wage and non-wage income in nominal terms, shown here, and then converting to real terms (shown below).
Non-wage income includes rents from property and other financial and non-financial assets, plus remittances. It is very difficult to model and so will likely be held as a fixed differential to wage income (i.e. if wage income increases by 2% then it is assumed that non-wage income increases by 2% as well).

(i18) RGDI = RRI – RDTX – REES + RWS + RBEN

| | |
|---|---|
| RGDI (reg) | nominal household incomes |
| RRI (reg) | non-wage income, nominal terms *(i92)* |
| RDTX (reg) | direct taxes, nominal terms *(i15)* |
| REES (reg) | employees' social contributions, nominal terms *(i16)* |
| RWS (reg) | wage income, nominal terms *(i57)* |
| RBEN (reg) | benefit payments, nominal terms *(i17)* |

### Calculating real incomes

Real incomes are determined by converting nominal incomes using the consumer price deflator.

(i19) RRPD = (RGDI / EX) / PRSC

| | |
|---|---|
| RRPD(reg) | measure of real household income, 2005 prices |
| RGDI (reg) | nominal household incomes *(i18)* |
| EX (reg) | exchange rate, convert national currency to euros *{x3}* |
| PRSC(reg) | aggregate consumer price index, 2005 = 1.0 *(i80)* |

## 1.8. E3ME Model: Econometric equations

[1] Disaggregated Final Consumption

CR = F (RRPD, PRCR, RRLR, PRSC/PSC1, CDEP, ODEP)

| | |
|---|---|
| CR(reg,con) | Final consumption by product |
| RRPD(reg) | disposable income, *(i19)* |
| PRCR(reg,con) | real prices, *(i41)* |
| RRLR(reg) | real long run interest rates, *(i45)* |
| PRSC(reg)/PSC1(reg) | change in price levels, *(i80)* |
| CDEP(reg), | child population dependency ratios *(i1)* |
| ODEP(reg) | aged population dependency ratios *(i2)* |

[2] Energy demand

FR0 = F (FRY, PREN, FRTD, ZRDM, ZRDT, FRK)
FRCT = F (FR0, PFRC, FRTD, ZRDM, ZRDT, FRK)
FRET = F (FR0, PFRE, FRTD, ZRDM, ZRDT, FRK)
FRGT = F (FR0, PFRG, FRTD, ZRDM, ZRDT, FRK)
FROT = F (FR0, PFRO, FRTD, ZRDM, ZRDT, FRK)

| | |
|---|---|
| FR0(reg,fu) | Total final energy demand |
| FRET(reg,fu) | Share of electricity demand |
| FRGT(reg,fu) | Share of natural gas demand |
| FROT(reg,fu) | Share of heavy oil demand |
| FRY(reg,fu) | Sectoral output, *(i71)* |
| PREN(reg,fu) | real average prices of energy, *(i42)* |
| FRTD(reg,fu) | R&D investment, *(i70)* |
| ZRDM(reg) | R&D expenditure for machinery, *(i68)* |
| ZRDT(reg) | R&D expenditure for vehicles, *(i67)* |
| FRK(reg,fu) | investment in energy using equipment *(i69)* |
| PFRC(reg,fu) | Price of coal €/toe *(i91)* |
| PFRE(reg,fu) | Price of electricity €/toe *(i90)* |
| PFRG(reg,fu) | Price of natural gas €/toe *(i91)* |
| PFRO(reg,fu) | Price of heavy oil €/toe *(i91)* |

[3] Investment into industrial sectors

KR = F (YR, PKR/PYR, YRWC, RRLR, YYN)

| | |
|---|---|
| KR(reg,inv) | Investment into industrial sectors |
| YR(reg,ind) | Industry output, *(i22)* |
| PKR(reg,ind)/PYR(reg,ind) | Investment prices to production ratio, *(i73) (i43)* |
| YRWC(reg,ind), | real labour costs *(i7)* |
| RRLR(reg) | real long run interest rates, *(i45)* |
| YYN(reg,ind) | production to production capacity ratio ('spare capacity') *(i49)* |

[4] Labour participation rate

LRP = F (RSQ, RWSR, RUNR, RBNR, RSER, RHRS, LRQU, RTIM)

| | |
|---|---|
| LRP(reg,lab) | Labour participation rate |
| RSQ(reg) | total output of products, *(i58)* |
| RWSR(reg) | purchasing power, *(i11)* |
| RUNR(reg) | unemployment rate, *(i13)* |
| RBNR(reg) | social security benefit to wages ratio, *{x10}* |
| RSER(reg) | ratio of services to non-services value added, *(i48)* |
| RHRS(reg) | hours worked, *(i56)* |
| LRQU(reg,lab) | qualifications, *{x4}* |
| RTIM(reg) | time trend, 1971=1, *{x17}* |

[5] Prices of exports

PQRX = F (PQRY, PQWE, EX, YRULT, YRKC*YRKS, YRKN)

| | |
|---|---|
| PQRX(reg,ind) | Prices of exports |
| PQRY(reg,ind) | prices of competing exports, *(i30)* |
| PQWE(reg,ind) | prices of world commodities, *(i74)* |
| EX(reg) | exchange rates, *{x3}* |

YRUL(reg,ind) +YRUT(reg,ind) unit labour costs plus unit tax costs, *(i35)*
YRKC(reg,ind)*YRKS(reg,ind) cumulated invest. IT tech times tertiary education skills, *(i83)* {x20}
YRKN(reg,ind) cumulated investment in non IT tech. *(i85)*

[6] Prices of imports,

PQRM = F (PQRF, PQWE, EX, YRUL, YRKC*YRKS, YRKN)
PQRM(reg,ind) Prices of imports
PQRF(reg,ind) weighted average of other regions' export prices, *(i31)*
PQWE(reg,ind) prices of world commodities, *(i74)*
EX(reg) exchange rates, {x3}
YRUL(reg,ind) +YRUT(reg,ind) unit labour costs plus unit tax costs, *(i35)*
YRKC(reg,ind)*YRKS(reg,ind) cumulated invest. IT tech times tertiary education skills, *(i83)* {x20}
YRKN(reg,ind) cumulated investment in non IT tech. *(i85)*

[7] Prices of domestic sales,

PYH = F (YRU, PQRM, YRKC*YRKS, YRKN, YYN)
PYH(reg,ind) Prices of domestic sales
YRUC(reg,ind) function of industrial unit costs, *(i34)*
PQRM(reg,ind) import prices, *[6]*
YRKC(reg,ind)*YRKS(reg,ind) cumulated invest. IT tech times tertiary education skills, *(i83)* {x20}
YRKN(reg,ind) cumulated investment in non IT tech. *(i85)*
YYN(reg,ind) production to production capacity ratio ('spare capacity') *(i49)*

[8] Product imports from outside trading region

QEM = F (QRDI, PQRM, PYH, EX, YRKC*YRKS, YRKN, SVIM, YYN)
QEM(reg,ind) Product imports from outside local trading region (e.g. EU)
QRDI(reg,ind) total supply, *(i24)*
PQRM(reg,ind) prices of imports, *[6]*
PYH(reg,ind) domestic prices, *[7]*
EX(reg) exchange rates, {x3}
YRKC(reg,ind)*YRKS(reg,ind) cumulated invest. IT tech times tertiary education skills, *(i83)* {x20}
YRKN(reg,ind) cumulated investment in non IT tech. *(i85)*
SVIM(reg) indicator of progress in the trade bloc {x18}
YYN(reg,ind) production to production capacity ratio ('spare capacity') *(i49)*

[9] Product imports from inside trading region

QIM = F (QRDI, PQRM, PYH, EX, YRKC*YRKS, YRKN, SVIM, YYN)
QIM(reg,ind) Product imports from inside local trading region
QRDI(reg,ind) total supply, *(i24)*
PQRM(reg,ind) prices of imports, *[6]*
PYH(reg,ind) domestic prices, *[7]*
EX(reg) exchange rates, {x3}
YRKC(reg,ind)*YRKS(reg,ind) cumulated invest. IT tech times tertiary education skills, *(i83)* {x20}
YRKN(reg,ind) cumulated investment in non IT tech. *(i85)*
SVIM(reg) indicator of progress in the trade bloc {x18}
YYN(reg,ind) production to production capacity ratio ('spare capacity') *(i49)*

[10] Investment in dwellings (houses)

RDW = F (RRPD, RRLR, CEDP, ODEP, RUNR, PRSC/PSC1)
RDW(reg) Investment in dwellings
RRPD(reg) disposable income, *(i19)*
RRLR(reg) real interest rates, *(i45)*
CDEP(reg), child population dependency ratios *(i1)*
ODEP(reg) aged population dependency ratios *(i2)*
RUNR(reg) unemployment, *(i13)*

PRSC(reg)/PSC1(reg) changes in price levels *(i80)*

[11] Total consumer expenditures

RSC = F (RRPD, RRLR, CDEP, ODEP, cum(RDW), RUNR, PRSC/PSC1)

| | |
|---|---|
| RSC(reg) | Total consumer expenditures, |
| RRPD(reg) | personal disposable income, *(i19)* |
| RRLR(reg) | real interest rates, *(i45)* |
| CDEP(reg), | child population dependency ratios *(i1)* |
| ODEP(reg) | aged population dependency ratios *(i2)* |
| cumulated RDW(reg) | cumulated investment in dwellings, *[10]* |
| RUNR(reg) | unemployment, *(i13)* |
| PRSC(reg)/PRC1(reg) | changes in price levels *(i80)* |

[12] Industry employment

YRE = F (YR, LYLC, YRH, YRKC*YRKS, YRKN)

| | |
|---|---|
| YRE(reg,ind) | Industry employment |
| YR(reg,ind) | industry output (equals demand QR(reg,ind)), *(i22)* |
| LYLC(reg,ind) | real wage costs per employee *(i6)* |
| YRH(reg,ind) | hours worked, *[13]* |
| YRKC(reg,ind)*YRKS(reg,ind) | cumulated invest. IT tech times tertiary education skills, *(i83)* *{x20}* |
| YRKN(reg,ind) | cumulated investment in non IT tech. *(i85)* |

[13] Hours worked

YRH = F (YRNH, YRKC*YRKS, YRKN, YYN)

| | |
|---|---|
| YRH(reg,ind) | Hours worked |
| YRNH(reg,ind) | standard hours worked in industry, *{x21}* |
| YRKC(reg,ind)*YRKS(reg,ind) | cumulated invest. IT tech times tertiary education skills, *(i83)* *{x20}* |
| YRKN(reg,ind) | cumulated investment in non IT tech. *(i85)* |
| YYN(reg,ind) | production to production capacity ratio ('spare capacity') *(i49)* |

[14] Industrial production capacity (Note: larger than YR = QR, i.e. there exists spare capacity)

YRN = F (YR9, YRKC*YRKS, YRKN, RWPP)

| | |
|---|---|
| YRN(reg,ind) | Industrial production capacity |
| YR9(reg,ind) | average growth in YR over the past nine years *(i22)* |
| YRKC(reg,ind)*YRKS(reg,ind) | cumulated invest. IT tech times tertiary education skills, *(i83)* *{x20}* |
| YRKN(reg,ind) | cumulated investment in non IT tech. *(i85)* |
| RWPP(reg) | working age population *(i53)* |

[15] Average earnings per person,

YRW = F (LYWE, LYRXE, LYRP, RUNR, RBNR, APSC, ARET, REIW, YYN)

| | |
|---|---|
| YRW(reg,ind) | Average earnings per person |
| YRWE(reg,ind) | wage rates in the same sector in other countries, *(i9)* |
| YRXE(reg,ind) | wage rates in other sectors in the same country *(i10)* |
| LYRP(reg,ind) | labour productivity, *(i5)* |
| RUNR(reg) | unemployment, *(i13)* |
| RBNR(reg), | social security benefits ratio to wages *{x10}* |
| APSC(reg) | consumer price deflator, adjusted for tax, *(i44)* |
| ARET(reg) | share of wages retained (not tax), *(i47)* |
| REIW(reg,ind) | inflation expectations, *{x13}* |
| YYN(reg,ind) | production to production capacity ratio ('spare capacity') *(i49)* |

[16] Long-term only econometric equations for bilateral trade

BIQRM(reg,ind,reg) = F (PQRX(reg,ind), YRKE(reg,ind))

PQRX(reg,ind) Prices of exports *[5]*
YRKE(reg,ind) Technology index *(i84)*

### 1.9. E3ME Model: Identity relationships

(NOTE: all variables possess the region (reg) classification when not specified)

Demographic factors

Child dependency ratio

(i1) CDEP= CPOP/RPOP *{x1}* *(i52)*

OAP dependency ratio

(i2) ODEP = OPOP/RPOP *(i52)* *{x6}*

The labour market and wages

Calculation of employees

(i3) YREE(ind) = fixed ratio of (YRE(ind)) *[12]*

Labour force is participation rates multiplied by population

(i4) LGR(lab) = LRP(lab) * DPAR(lab) *[4]* *{x2}*

Labour productivity (in current prices)

(i5) LYRP(ind) = PYR(ind) * YR(ind) / YRE(ind) *(i43)* *(i22)* *[12]*

Labour costs per employee in real terms

(i6) LYLC(ind) = (YRLC(ind) / PYR(ind)) / YREE(ind) *(i8)* *(i43)* *(i3)*

Average wage costs in real terms

(i7) YRWC(ind) = (YRLC(ind) / PYR(ind)) / YREE(ind) *(i8)* *(i43)* *(i3)*

Labour costs, consisting of wages and social security payments

(i8) YRLC(ind) = YRWS(ind) * (1 + RERR) *(i14)* *{x13}*

Wage rates in other sectors (same region)

(i9) YRWE(ind) = sum(LN(YRW(i)) * YRLC(i) / sum(YRLC(i)) - LN(APSC)) *(for i = all other industries)* *[15]* *(i8)* *(i44)*

Wage rates in other regions (same sector)

(i10) YRXE(ind) = sum(LN(YRW(j)) * YRLC(j) / sum(YRLC(j)) - LN(APSC)) *(for j = all other countries)* *[15]* *(i8)* *(i44)*

Average wage rates per employee, real terms

(i11) RWSR = EX * RWS / (PRSC * REEM) *(i80)* *{x3}* *(i57)* *(i55)*

Unemployment, in thousands of people

(i12) RUNE = RWPP – REMP *(i53)* *(i54)*

Unemployment rate

(i13) RUNR = RUNE / RWPP *(i12)* *(i53)*

Determining household income

Total sectoral wages

(i14) YRWS(ind) = YREE(ind) * YRW(ind) *(i3)* *[15]* *(i22)*

Income tax receipts (rate * wages)

(i15) RDTX = RDTR * RWS *(i57)* *{x11}*

Employee social security receipts

(i16) REES = REER * RWS *(i57)* *{x12}*

Benefit payments

(i17) RBEN = RBNR * RWS *{x10}* *(i57)*

Nominal household income = wages + other income + benefits, minus taxes

(i18) RGDI = RRI – RDTX – REES + RWS + RBEN *(i92)* *(i15)* *(i16)* *(i57)* *(i17)*

Household incomes in real terms

(i19) RRPD = (RGDI*EX/PRSC) *(i80)* *{x3}* *(i18)*

Rate of retained wages (i.e. untaxed)

(i20) RRET = RWS / (RWS – RDTX – REES) *(i57)* *(i15)* *(i16)*

Non-wage income

(i92) RRI = fixed ratio of RWS *(i57)*

Calculating GVA and GDP

GDP identity

RGDP = RSC + RSK + RSG + RSX – RSM *(i59)* *(i60)* *(i61)* *(i62)* *(i63)*

Identity for (gross) output, summing demand components and subtracting imports

(i21) QR(ind) = QRY(ind) + QRC(ind) + QRK(ind) + QRG(ind) + QRX(ind) – QRM(ind) *(i28)* *(i29)* *(i75)* *(i76)* *(i78)* *(i77)*

Assumption that supply adjusts to meet demand

(i22) YR(ind) = QR(ind) *(i21)*

GVA at factor cost, determined by output minus costs (incl product taxes)

YRF(ind) = YR(ind) * (1 – YRUM(ind) – YRUT(ind)) *(i22)* *(i34)* *(i37)*

GVA at market prices, determined by output minus costs

(i23) YRM(ind) = QR(ind) * (1 – YRUC(ind)) *(i21)* *(i34)*

Total supply, domestic production plus imports

(i24) QRDI(ind) = QR(ind) + QRM(ind) *(i21)* *(i28)*

Output in current prices

(i25) VQR(ind) = PQR(ind) * QR(ind) *(i40)* *(i21)*

Modelling trade

Exports from outside local trading region – determined by reversing bilateral import flows

(i26) QEX(ind) = reverse (BIQRM) *[16]*

Exports from inside local trading region – determined by reversing bilateral import flows

(i27) QIX(ind) = reverse (BIQRM) *[16]*

Total imports, both within and outside local trading region

(i28) QRM(ind) = QIM(ind) + QEM(ind) *[9]* *[8]*

Total exports, both within and outside local trading region

(i29) QRX(ind) = QIX(ind) + QEX(ind) *(i26)* *(i27)*

Price of competing goods (weighted average of other regions' exports)

(i30) PQRY(ind) = sum (QZXC(ind,reg)*VQRX(ind,reg)) / SUM(QZXC(ind,reg)*QRX(ind,reg)) *(i32) (i29)*

Price of competing goods (weighted average of other regions' imports)

(i31) PQRF(ind) = SUM(QZMC(ind,reg)) * VQRX(ind,reg)) / SUM(QZMC(ind,reg) * QRX(ind,reg)) *(i32) (i29) {x9}*

Exports in current prices

(i32) VQRX(ind) = PQRX(ind) * QRX(ind) *[5] (i29)*

Imports in current prices

(i33) VQRM(ind) = PQRM(ind) * QRM(ind) *[6] (i28)*

Sectoral unit costs

Total unit costs, made up of material, labour and taxation costs

(i34) YRUC(ind) = YRUM(ind) + YRUL(ind) + YRUT(ind) *(i36) (i35) (i37)*

Unit labour costs

(i35) YRUL(ind) = YRLC(ind) / YR(ind) *(i8) (i22)*

Unit material costs, determined by IO coefficients

(i36) YRUM(ind) = sum (QYC(ind) * PQRD(ind)) / YR(ind) *(i22) (i39) {x8}*

Unit taxation costs

(i37) YRUT(ind) = YRT(ind) / YR(ind) *(i46) (i22)*

Price formation

Commodity prices in euros (converted from USD)

(i38) PM(com) = PMF(com) * EX(34, t = 2005) / EX(reg = USA) *(i39) {x7} {x3}*

Prices of goods for domestic consumption

(i39) PQRD(ind) = (VQR(ind) + VQRM(ind) – VQRX(ind)) / (QR(ind) + QRM(ind) – QRX(ind)) *(i25) (i33) (i32) (i21) (i28) (i29)*

Product prices (= industry prices in E3ME)

(i40) PQR(ind) = PYR(ind) *(i43)*

Relative prices of consumption goods

(i41) PRCR(con) = PCR(con) / PRSC *(i80) (i72)*

Fuel prices in real terms

(i42) PREN(FU) = PFR0(FU) / PRSQ *(i79) (i81)*

Industry prices, based on product prices (simplified representation)

(i43) PYR(ind) = PQRD(ind) (this is a slight simplification) *(i39)*

Consumer prices adjusted for changes in taxation rates

(i44) APSC = PRSC + RRET *(i80) (i20)*

Real interest rate

(i45) RRLR = 1 + (RLR–LN(PRSC/PSC1)) / 100 *(i80) {x16}*

Taxes paid by each industry

(i46) YRT(ind) = YRTR(ind) * YR(ind) *{x22} (i22)*

Miscellaneous identity relationships

(i47) ARET = RRET * (1 + RERR) * RITR *(i20)* *{x13}* *{x15}*

Share of services in the economy

(i48) RSER = RSERV / NSERV *(i65)* *(i66)*

Ratio of output to normal output

(i49) YYN(ind) = YR(ind) / YRN(ind) *(i22)* *[14]*

Demographic factors

Total population

(i52) RPOP = sum(DPAR (lab)) *{x2}*

Working age population

(i53) RWPP = sum (LGR(lab)) *(i4)*

The labour market and household incomes

Total employment

(i54) REMP = sum (YRE(ind)) *[12]*

Total number of employees

(i55) REEM = sum (YREE) *(i3)*

Total working hours

(i56) RHRS = sum (YRH(ind) * YRE(ind)) *[12]* *[13]*

Total wages

(i57) RWS = sum(YRWS(ind)) *(i14)*

Calculating GVA and GDP

Total output

(i58) RSQ = SUM(QR) *(i21)*

Total consumption, investment, government spending, exports and imports

(i59) RSC = sum(CR) *[1]* *[11]* *(NOTE: RSC and CR are both econometric; this identity is used to ensure that both variables are consistent with an unallocated consumption category in CR)*

(i60) RSK = sum(KR) *[3]*

(i61) RSG = sum(GR) *{x5}*

(i62) RSX = sum(QRX) *(i29)*

(i63) RSM = sum(QRM) *(i28)*

Total GVA, current prices

(i64) VRYM = sum (YRM(ind) * PYRM(ind))

Services and non-services production

Total services production

(i65) RSERV = sum (YR(i)) (for i = all service industries) *(i22)*

Level of non-service activity in the economy

(i66) NSERV = sum (YR(i)) (for i = all non-service industries) *(i22)*

Global R&D for spillover effects

Global R&D in vehicles sector

(i67) ZRDT = sum(YRDS) for motor vehicle industry globally *{x19}*

Global R&D in machinery sector

(i68) ZRDM = sum(YRDS) for machinery industries globally *{x19}*

Classification converters

(i69) FRK(FU) = FUYC * KR(ind) *⟨1⟩ [3]*

(i70) FRTD(FU) = FUYC*YRDS(ind) *⟨1⟩ {x19}*

(i71) FRY(FU) = FUYC * YR(ind) *⟨1⟩ (i69)*

(i72) PCR(con) = QCC * PQRD(ind) * (1+CRTR) *⟨3⟩ (i39)* {x25}

(i73) PKR(inv) = QKC * PQRD(ind) *⟨4⟩ (i39)*

(i74) PQWE(ind) = QMC* PM(com) *⟨6⟩ (i38)*

(i75) QRY(ind) = QYC(ind) * YR(ind) *(matrix calculation across all industries) (i22) {x8}*

(i76) QRC(ind) = QCC * CR(con) *⟨3⟩ [1]*

(i77) QRG(ind) = QGC * GR(gov) *⟨5⟩ {x5}*

(i78) QRK(ind) = QKC * KR(inv) *⟨4⟩ [3]*

Weighted averages

(i79) PFR0(FU) = ave(PFRC(FU),PFR2(FU),PFR3(FU),PFRO(FU),PFRM(FU),PFR6(FU),PFRG(FU),PFRE(FU),PFR9(FU),PF10(FU),PF11(FU),PF12(FU)), weighted by FRCT(FU),FR02(FU),FR03(FU),FROT(FU),FR05(FU),FR06(FU),FRGT(FU),FRET(FU),FR09(FU), FR10(FU),FR11(FU),FR12(FU), respectively *[2] (i86) (i87) (i90) (i91)*

(i80) PRSC = ave (PCR(con)), weighted by CR(con) *(i72) [1]*

(i81) PRSQ = ave (PQR(ind)), weighted by QR(ind) *(i21) (i40)*

Technology indices

*These use a separate accumulation function. They are split to cover ICT investment (YRKC), and other investment (YRKN). For regions that do not have separate data, total investment is used (YRKE).*

YRKC at time t is written as:

$$YRKC_t = c + \alpha dt(\tau 1)$$

where $dt(\tau 1)$ satisfies the following recursive formula

$$dt(\tau 1) = \tau 1 * d_{t-1}(\tau 1) + (1 - \tau 1) * \log(KR_t + \tau 2 * YRDS_t)$$

In the current version of E3ME, based on a calibration exercise:

$\tau 1 = 0.3$

$\tau 1 = 5.0$

(i83) YRKC(ind) = F ($KR_{ICT}$(ind), YRDS(ind)) *[3]* *{x19}*

(i84) YRKE(ind) = F ($KR_{all}$(ind), YRDS(ind)) *[3]* *{x19}*

(i85) YRKN(ind) = F ($KR_{non\text{-}ICT}$(ind), YRDS(ind)) *[3]* *{x19}*

FTT feedbacks

Fuel use for power, transport, households (FU = 1, 16, 19)

(i86) FRCT(FU=1), FROT(FU=1), FRGT(FU=1) = MJEF(ft) *(fi8)*

(i87) FRET(FU=16), FR05(FU=16), FRGT(FU=16), FRBT(FU=16) = TJEF(ft) *(fi18)*

Investment in electricity generation equipment

(i89) KR(inv=26 for EU, 22 for not EU) scaled by % changes in sum(MWIY(ptech)) *[3]*

Marginal costs of production for electricity, oil, coal, gas, for determining fuel prices

(i90) PFRE determined by MEWP(ft=8) plus taxes *(fi10)*

(i91) PFRO, PFRC, PFRG, PFRM historical value scaled by changes in MERC(eres) plus energy taxes *[f5]*

Exogenous variables

{x1} CPOP: child population

{x2} DPAR: Population split into age band and gender

{x3} EX: exchange rates, local currency per euro, 2005=1

{x4} LRQU: (logged) qualifications mix for 27 age/gender groups

{x5} GR: Government spending in real terms by broad category

{x6} OPOP: old-age population

{x7} PMF: global commodity prices

{x8} QYC: input-output coefficient matrix

{x9} QZMC: bilateral trade shares of industry imports by origin

{x10} RBNR: social benefit rates, as a proportion of wages

{x11} RDTR: Income tax rates

{x12} REER: Employee social security rates

{x13} REIW: Assumption about inflation expectations, e.g. determined by central bank target (%)

{x14} RERR: Employer social security rates

{x15} RITR: Indirect tax rates

{x16} RLR: long-run nominal interest rates

{x17} RTIM: time trend

{x18} SVIM: an indicator of progress in the trade bloc

{x19} YRDS: R&D expenditure by industry

{x20} YRKS: skills matrix, taken from last year of data

{x21} YRNH: normal hours worked per week

{x22} YRTR: tax rates by industry (taken from last year of data)

{x23} REPP: Carbon price €/tC

{x24} RTEA: Energy tax €/toe

{x25} CRTR: Tax rates on consumer products (rate)

Converter matrices

⟨1⟩ FUYC: ind → FU

⟨2⟩ QYC: Product→ ind (this is the input-output table)

⟨3⟩ QCC: con → ind

⟨4⟩ QKC: inv → ind (the sector making the investment, to the one that builds the products)

⟨5⟩ QGC: Gov → ind

⟨6⟩ QMC: Commodities → Ind

Dimensions

com: Commodity

con: Consumption product

fu: Fuel user

gov: Govt expenditure category

ind: Industry

inv: Investing industry

reg: Region

lab: Labour force group (gender/age)

ft: Fuel type

ptech: FTT:Power technology types

ttech: FTT:Transport technology types

htech: FTT:Heat technology types

eres: FTT energy resource types

### 1.10. FTT:Power equations

[f1] FTT:Power technology market shares (non-linear differential equation)

$\Delta MEWS = MEWS \times sum(MEWS \times [F(MEWC,MEWD) - F^*(MEWC,MEWD)])\ \Delta t$

| | |
|---|---|
| MEWS(reg,ptech) | Shares of power technology |
| ΔMEWS(reg,ptech) | Change in shares over a unit of time Δt (quarterly) |
| Fij | Discrete choice binary preference matrix *(fi1)* |
| Fji | Transpose of discrete choice binary preference matrix *(fi1)* |

For more information on the model structure and calculation, see Mercure (2012), Mercure et al (2014), Mercure (2015). In particular, see Mercure (2012) for an explanation of the treatment of renewables intermittency, capacity constraints and endogenous capacity factors.

Main FTT:Power Identity relationships

Discrete choice model preferences matrix (element i,j)

(fi1) $F_{ij} = 0.5 \times [1+\tanh(1.25 \times (MEWC(i) - MEWC(j))/sqrt(MEWD(i)^2 + MEWD(i)^2)]$

NOTE: only if MREG = 1, otherwise Fij = 0 and Fji = 1

| | |
|---|---|
| MEWC(reg,ptech) | Mean power technology costs, \$/MWh *(fi4)* |
| MECD(reg,ptech) | Standard deviation power technology costs, \$/MWh *(fi5)* |
| MREG(reg,ptech) | Technology regulation (whether something can be built) *{fx4}* |

Note: the 0.5×(1+tanh(1.25×[...])) function is the logistic curve of the logit model. Each element of the matrix originates from a pair-wise cost comparison binary logit, and Fij + Fji = 1.

Power generation capacity from shares and demand

(fi2) MEWK = MEWS*MEWD/sum(MEWS×MEWL)
MEWK(reg,ptech) Electricity capacity in GW
MEWS(reg,ptech) Shares of power technology *[f1]*
MEWD(reg) Total electricity generation (= FRET + losses) GWh/y *(fi7)*
MEWL(reg,ptech) Load factor *{fx1}* (See Mercure (2012) endogenous treatment)

Power generation from capacity and load factors

(fi3) MEWG = MEWK×MEWL*8766
MEWG(reg,ptech) Electricity generation in GWh/y
MEWK(reg,ptech) Electricity capacity in GW *(fi2)*
MEWL(reg,ptech) Load factor *{fx1}* (See Mercure (2012) endogenous treatment)

Mean levelised cost for generating electricity by technology

(fi4) $MEWC = \Sigma[(INV(t) \times SUB(t)/CF(t) + FUEL(t) + OM(t) + TAX(t))/(1+r)^t]/\Sigma[1/(1+R)^t]$ - FiT
MEWC(reg,ptech) Levelised cost for generating electricity including taxes, \$/MWh
INV(reg,ptech) Power technology investment costs in \$/KW *{fx1}* *(fi36)*
CF(reg,ptech) Expected capacity factor *{fx1}* *(fi36)*
FUEL(reg,ptech) Power technology fuel costs in \$/MWh *{fx1}* *[f5]* *(fi36)*
OM(reg,ptech) Power technology operation & maintenance costs in \$/MWh *{fx1}*
TAX(reg,ptech) Carbon price or tax converted to \$/MWh *{x23}*
SUB(reg,ptech) Capital cost subsidy *{fx2}*
FiT(reg,ptech) Feed-in Tariff subsidy *{fx3}*

STD levelised cost for generating electricity by technology

(fi5) $MECD = \Sigma[\mathrm{sqrt}(\Delta INV(t)^2 + \Delta FUEL(t)^2 + \Delta OM(t)^2)/(1+r)^t]/\Sigma[1/(1+R)^t]$
ΔINV(reg,ptech) STD Power technology investment costs in \$/KW *{fx1}*
ΔFUEL(reg,ptech) STD Power technology fuel costs in \$/MWh *{fx1}*
ΔOM(reg,ptech) STD Power technology operation & maintenance in \$/MWh *{fx1}*

Learning-by-doing curves

(fi6) $\Delta INV = - LR \times INV \times \Delta MEWW/MEWW$
INV(reg,ptech) STD Power technology investment costs in \$/KW *{fx1}*
LR(ptech) Learning power law exponents (global) *{fx1}*
MEWW(ptech) Global cumulative installed capacity in GW *(fi9)*

Total electricity generation

(fi7) MEWD = sum(FRET) + MELO
MEWD(reg) Total electricity generation, GWh/y
FRET(reg,fu) E3ME electricity demand, th toe *[2]*
MELO(reg) Losses (IEA data, equal to generation – demand, constant) *{fx1}*
Note: MEWD = sum(MEWG) by definition

Fuel use for power generation

(fi8) MJEF(reg,ft) = sum(MEWG×MMEF)
MEWG(reg,ptech) Electricity generation by technology *(fi3)*
MMEF(ptech,ft) IEA/IPCC Fuel use factors *{fx1}*

Global Cumulative capacity

(fi9) MEWW = { sum(ΔMEWK/Δτ, τ = 1970 to t) + MEWK/ LT, ΔMEWK/Δτ > 0
MEWK/ LT ΔMEWK/Δτ < 0

MEWW(ptech) Global cumulative installed capacity in GW
MEWK(reg,ptech) Electricity capacity in GW *(fi2)*
LT(ptech) Technology lifetimes *{fx1}*

Marginal cost of electricity generation (weighted average of the levelised cost)

(fi10) MEWP = sum(MEWC*MEWG)/MEWD
MEWP(reg) Marginal cost of electricity generation ($/MWh)
MEWG(reg,ptech) Electricity generation by technology *(fi3)*
MEWC(reg,ptech) Levelised cost for generating electricity including taxes, $/MWh *(fi4)*
MEWD(reg) Total electricity generation, GWh/y *(fi7)*

FTT:Power policy inputs and exogenous values

{fx1} BETC(reg,ptech): Technology characteristics (capital, fuel, O&M, lifetimes, load factors etc)

{fx2} MEWT(reg,ptech): Capital cost subsidies (in % paid by the gov.)

{fx3} MEFI(reg,ptech): Feed-in tariffs (as % of LCOE difference to electricity price paid by the grid)

{fx4} MREG(reg,ptech): Regulations (whether a technology can be built or not)

### *1.11. FTT:Transport equations*

[f2] FTT:Transport technology market shares (non-linear differential equation)

ΔTEWS = TEWS×Σ(TEWS×[F(TEWC,TEWD) – F*(TEWC,TEWD)]) Δt
TEWS(reg,ttech) Shares of transport technology capacity
ΔTEWS(reg,ttech) Change in shares over a unit of time Δt (quarterly)
TEWC(reg,ttech) Mean transport technology costs, $/person-kilometre *(fi15)*
TECD(reg,ttech) Standard deviation transport technology costs, $/p-km *(fi16)*
F(TEWC,TEWD) Discrete choice binary preference matrix *(fi10)*
F*(TEWC,TEWD) Transpose of discrete choice binary preference matrix *(fi10)*

[f3] Demand for new private passenger vehicles

RVEH = F(RRPD, RPVE)
RVEH(reg) Total new vehicle sales, in th veh
RRPD(reg) Disposable income, *(i19)*
RPVE(reg) Average vehicle prices, *{fx11}*

[f4] Demand for transport services

RVKM = F(RRPD, RPVE)
RVKM(reg) Demand for transport services, Mvkm/y
RRPD(reg) Disposable income, *(i19)*
PFRM(reg,ft = middle distillates) Fuel prices, *(i91)*

Main FTT:Transport Identity relationships

Discrete choice model preferences matrix (element i,j)

(fi10) Fij(TEWC,TEWD) = 0.5×[1+tanh(1.25×(TEWC(i) – TEWC(j))/sqrt(TEWD(i)$^2$+ TEWD(i)$^2$)]
NOTE: only if TREG = 1, otherwise Fij = 0 and Fji = 1
TEWC(reg,ttech) Mean perceived vehicle technology costs, $/pkm *(fi4)*
TECD(reg,ttech) Standard deviation vehicle technology costs, $/ pkm *(fi5)*
TREG(reg,ptech) Technology regulation (whether a type of vehicle can be sold) *{fx9}*
Note: the 0.5×(1+tanh(1.25×[...])) function is the logistic curve of the logit model. Each element of the matrix originates from a pair-wise cost comparison binary logit, and Fij + Fji = 1.

Fleet size: survival analysis

(fi11) RFLT = sum(RVEH*TESF, past 20 years up to the present)

| | |
|---|---|
| RFLT(reg) | Vehicle fleet size, in th veh |
| RVEH(reg) | Total new vehicle sales (historical + modelled), in th veh *[f3]* |
| TESF(reg) | Vehicle survival function *{fx5}* |

Transport fleet capacity from shares and demand

(fi12) TEWK = TEWS*RFLT

| | |
|---|---|
| TEWK(reg,ttech) | Vehicle fleet capacity per technology type (k-seats) |
| TEWS(reg,ttech) | Shares of transport technology capacity *[f1]* |
| RFLT(reg) | Total transport fleet size, th veh *(fi11)* |

Transport fleet utilisation rate (Mpkm/y/k-veh)

(fi13) TEWL = RVKM/RFLT*BTTC

| | |
|---|---|
| TEWL(reg,ttech) | Transport fleet use factors |
| RVKM(reg) | Demand for transport services, Mvkm/y *[f4]* |
| RFLT(reg) | Total transport fleet size, th veh *(fi11)* |
| BTTC(reg) | Vehicle use parameters (distance, occupancy) *{fx5}* |

Transport generation from capacity and load factors

(fi14) TEWG = TEWK×TEWL

| | |
|---|---|
| TEWG(reg,ttech) | Transport service generation in Mpkm/y |
| TEWL(reg,ttech) | Transport fleet use factors *(fi13)* |
| TEWK(reg,ttech) | Vehicle fleet capacity per technology type (k-seats) *(fi12)* |

Mean levelised cost for transport by technology

(fi15) $TEWC = \Sigma[(INV(t) \times REG(t) + FUEL(t) + FTAX(t) + OM(t) + RTAX(t))/(1+r)^t]/\Sigma[1/(1+R)^t]$

| | |
|---|---|
| TEWC(reg, ttech) | Levelised perceived cost of generating transport services, $/pkm |
| INV(reg, ttech) | Vehicle prices per expected distance driven $/pkm *{fx5}* |
| FUEL(reg, ttech) | Vehicle fuel costs in $/pkm *{fx5}* |
| OM(reg, ttech) | Operation & maintenance costs in $/MWh *{fx5}* |
| FTAX(reg, ttech) | Fuel Taxes *{fx7}* |
| RTAX(reg, ttech) | Road Taxes *{fx8}* |
| REG(reg, ttech) | Vehicle registration taxes *{fx6}* |

STD levelised cost for generating electricity by technology

(fi16) $TECD = \Sigma[\mathrm{sqrt}(INV(t)^2 + FUEL(t)^2 + OM(t)^2)/(1+r)^t]/\Sigma[1/(1+R)^t]$

| | |
|---|---|
| TEWC(reg, ttech) | Levelised perceived cost of generating transport services, $/pkm |
| ΔINV(reg, ttech) | Vehicle prices per expected distance driven $/pkm *{fx5}* |
| ΔFUEL(reg, ttech) | Vehicle fuel costs in $/pkm *{fx5}* |
| ΔOM(reg, ttech) | Operation & maintenance costs in $/pkm *{fx5}* |

Learning-by-doing curves

(fi17) $\Delta INV = - LR \times INV \times \Delta TEWW/TEWW$

| | |
|---|---|
| INV(reg, ttech) | Vehicle prices per expected distance driven $/pkm *{fx5}* |
| LR(ttech) | Learning power law exponents (global) *{fx5}* |
| TEWW(ttech) | Global cumulative vehicle sales per tech type in k-veh *(fi19)* |

Fuel use for transport

(fi18) TJEF(reg) = sum(TEWG ×TTEF(1-RBFM))

NOTE: only for liquid fuel-using technologies

| | |
|---|---|
| TEWG(reg,ttech) | Transport service generation in Mpkm/y *(fi14)* |
| TTEF(ttech,ft) | Vehicle fuel economy factors *{fx5}* |
| RBFM(reg) | Biofuel mandate (% of liquid fuel served at the pump) *{fx10}* |

Global Cumulative capacity

(fi19) $TEWW = \begin{cases} sum(\Delta TEWK/\Delta\tau,\ \tau = 1970 \text{ to } t) + TEWK/LT, & \Delta TEWK/\Delta\tau > 0 \\ TEWK/LT & \Delta TEWK/\Delta\tau < 0 \end{cases}$

| | |
|---|---|
| TEWW(ttech) | Global cumulative vehicle sales per tech type in k-veh |
| TEWK(reg,ttech) | Vehicle fleet capacity per technology type (k-seats) *(fi12)* |
| LT(ttech) | Technology lifetimes *{fx5}* |

FTT:Transport policy inputs and exogenous values

{fx5} BTTC(reg, ttech): Technology characteristics (capital, fuel, O&M, lifetimes, load factors etc)

{fx6} RCO2(reg), TTVT(reg, ttech): Registration tax / fuel economy tax (in \$/veh or \$/g$CO_2$/km)

{fx7} RTFT(reg): Fuel tax (\$/litre)

{fx8} TTRT(reg,ttech): Road tax (\$/year)

{fx9} TREG(reg,ttech): Regulations (whether a type of vehicle can be sold)

{fx10} RBFM(reg): Biofuel mandate (% of liquid fuel served at the pump)

{fx11} RPVE(reg): Average vehicle prices

### 1.1. FTT:Heat equations

[f5] FTT:Heat technology market shares (non-linear differential equation)

$\Delta HEWS = HEWS \times \Sigma(HEWS \times A([Fij/LT - Fji/LT] + [FEij/PB - FEji/PB]))\ \Delta t$

| | |
|---|---|
| HEWS(reg,htech) | Shares of heating technology in total heat generation |
| ΔHEWS(reg,htech) | Change in shares over a unit of time Δt (quarterly) |
| A(htech) | Matrix of exogenous behavioural assumptions |
| LT(reg,htech) | Technology lifetimes |
| HGC1(reg,htech) | Mean heating technology costs, \$/kWh of heat produced *(fi24)* |
| HGD1(reg,htech) | STD heating levelised costs, \$/kWh of heat produced *(fi27)* |
| FE(HGC1, HGD1) | Discrete choice binary preference matrix, regular choices *(fi20)* |
| FE*(HGC1, HGD1) | Transpose of binary preference matrix, regular choices *(fi20)* |
| FE(HGC2,HGD2, HGC3,HGD3) | Binary preference matrix, premature replacement choices *(fi20)* |
| FE*( HGC2,HGD2, HGC3,HGD3) | Transpose of binary preference, premature replacement *(fi21)* |
| PB(reg,htech) | Behavioural payback-time threshold in years (fx12) |

Discrete choice model preferences matrix – regular choices (element i,j)

(fi20) $Fij(HGC1,HGD1) = 0.5 \times [1+\tanh(1.25 \times (HGC1(i) - HGC1(j))/sqrt(HGD1(i)^2 + HGD1(i)^2)]$

NOTE: only if HREG = 1, otherwise Fij = 0 and Fji = 1

| | |
|---|---|
| HGC1(reg,htech) | Mean perceived heating levelised costs, \$/kWh *(fi24)* |
| HGD1(reg,htech) | Standard deviation heating levelised costs, \$/kWh *(fi27)* |
| HREG(reg,htech) | Technology regulation (whether a type of heating system can be sold) *{fx15}* |

Note: the 0.5×(1+tanh(1.25×[...])) function is the logistic curve of the logit model. Each element of the matrix originates from a pair-wise cost comparison binary logit, and Fij + Fji = 1.

Discrete choice model preferences matrix – premature replacement choices (element i,j)

(fi21) $FEij(HGC2,HGD2,HGC3,HGD3) = 0.5\times[1+\tanh(1.25\times(HGC2(i) - HGC3(j))/\mathrm{sqrt}(HGD2(i)^2+HGD3(i)^2)]$

$FEji(HGC2,HGD2,HGC3,HGD3) = 0.5\times[1+\tanh(1.25\times(HGC2(j) - HGC3(i))/\mathrm{sqrt}(HGD2(j)^2+HGD3(i)^2)]$

NOTE: only if HREG = 1, otherwise FEij = 0 and Gji = 1

| | |
|---|---|
| HGC2(reg,htech) | Mean perceived heating marginal costs, $/kWh *(fi25)* |
| HGD2(reg,htech) | Standard deviation heating marginal costs, $/kWh *(fi28)* |
| HGC3(reg,htech) | Mean perceived heating payback-based costs, $/kWh *(fi26)* |
| HGD3(reg,htech) | Standard deviation heating payback-based costs, $/kWh *(fi29)* |
| HREG(reg,htech) | Technology regulation (if a type of heating system can be sold) *{fx9}* |

Heat generation from shares and demand

(fi22) HEWG = HEWS*RHUD

| | |
|---|---|
| HEWG(reg,htech) | Heat generation in GWh/y |
| HEWS(reg,htech) | Shares of heating technology *[f5]* |
| RHUD(reg) | Total heat generation GWh/y *(fx17)* |

Heating capacities

(fi23) HEWK = HEWG*CF

| | |
|---|---|
| HEWK(reg,htech) | Heating capacity in GW |
| HEWG(reg,htech) | Heat generation in GWh/y (fi22) |
| CF(reg,htech) | Capacity factor (climate- and technology-dependent) (fx12) |

Mean levelised cost for generating heat by technology

(fi24) $HGC1 = \Sigma[(INV(t)\times SUB(t)/CF(t) + FUEL(t) + OM(t) + TAX(t) - FiT(t))/(1+ r)^t]/\Sigma[1/(1+ r)^t]$

| | |
|---|---|
| HGC1(reg,htech) | Levelised perceived cost for generating heat including taxes, $/kWh |
| INV(reg,htech) | Heating technology investment costs in $/kW *{fx12}* *(fi30)* |
| CF(reg,htech) | Expected capacity factor *{fx12}* |
| FUEL(reg,htech) | Heating technology fuel costs in $/kWh of heat *{fx12}* |
| OM(reg,htech) | Heating technology operation & maintenance costs in $/kWh *{fx12}* |
| TAX(reg,htech) | Carbon price or tax converted to $/kWh *{fx16}* |
| SUB(reg,htech) | Capital cost subsidy *{fx13}* |
| FiT(reg,htech) | Feed-in Tariff subsidy *{fx14}* |

Mean marginal cost for generating heat by technology

(fi25) $HGC2 = FUEL(t) + OM(t) + TAX(t) - FiT(t)$

| | |
|---|---|
| HGC2(reg,htech) | Marginal perceived cost for generating heat incl. taxes, $/kWh |
| FUEL(reg,htech) | Heating technology fuel costs in $/kWh of heat *{fx12}* |
| OM(reg,htech) | Heating technology operation & maintenance costs in $/kWh *{fx12}* |
| TAX(reg,htech) | Carbon price or tax converted to $/kWh *{fx16}* |
| FiT(reg,htech) | Feed-in Tariff subsidy *{fx14}* |

Mean payback-based cost for generating heat by technology

(fi26) $HGC3 = HGC2 + (INV(t)\times SUB(t)/CF(t))/PB$

| | |
|---|---|
| HGC3(reg,htech) | Payback-based perceived cost for generating heat incl. taxes, $/kWh |
| INV(reg,htech) | Heating technology investment costs in $/kW *{fx12}* *(fi30)* |
| CF(reg,htech) | Expected capacity factor *{fx12}* |
| SUB(reg,htech) | Capital cost subsidy *{fx13}* |
| PB(reg,htech) | Behavioural payback-time threshold in years *{fx12}* |

STD levelised cost for generating heat by technology

(fi27) $HGD1 = \Sigma[\mathrm{sqrt}(\Delta INV(t)^2 + \Delta FUEL(t)^2 + \Delta OM(t)^2)/(1+ r)^t]/\Sigma[1/(1+ r)^t]$

| | |
|---|---|
| ΔINV(reg,htech) | STD Heating technology investment costs in $/kW *{fx12}* *(fi30)* |
| ΔFUEL(reg,htech) | STD Heating technology fuel costs in $/kWh *{fx12}* |

$\Delta$OM(reg,htech) STD Heating technology operation & maintenance in \$/kWh *{fx12}*

STD marginal cost for generating heat by technology

(fi28) HGD2 = sqrt($\Delta$FUEL(t)$^2$ + $\Delta$OM(t)$^2$)

$\Delta$FUEL(reg,htech) STD Heating technology fuel costs in \$/kWh *{fx12}*

$\Delta$OM(reg,htech) STD Heating technology operation & maintenance in \$/kWh *{fx12}*

STD payback-based cost for generating heat by technology

(fi29) HGD3 = sqrt($\Delta$FUEL(t)$^2$ + $\Delta$OM(t)$^2$ + $\Delta$INV(t)$^2$/PB$^2$ + $\Delta$PB$^2$*(INV(t)$^2$/PB$^4$))

$\Delta$INV(reg,htech) STD Heating technology investment costs in \$/kW *{fx12}*

$\Delta$FUEL(reg,htech) STD Heating technology fuel costs in \$/kWh *{fx12}*

$\Delta$OM(reg,htech) STD Heating technology operation & maintenance in \$/kWh *{fx12}*

PB(reg,htech) Behavioural payback-time threshold in years

Learning-by-doing curves

(fi30) $\Delta$INV = - LR$\times$INV$\times\Delta$HEWW/HEWW

INV(reg,htech) Heating technology investment costs in \$/kW *{fx12}*

LR(htech) Learning power law exponents (global) *{fx12}*

HEWW(htech) Global cumulative installed capacity in GW *(fi33)*

Final energy demand for heat generation

(fi31) HEWF(reg,htech) = HEWG$\times$CE

HEWG(reg,htech) Heat generation by technology *(fi22)*

CE(reg,htech) Conversion efficiency by technology *{fx12}*

Fuel use for heat generation

(fi32) HJHF(reg,ft) = sum(HEWF$\times$HJET)

HEWG(reg,htech) Heat generation by technology *(fi22)*

HJET(htech,ft) Fuel use factors *{fx12}*

Global Cumulative capacity

(fi33) HEWW = { sum($\Delta$HEWK/$\Delta\tau$, $\tau$ = 1990 to t) + HEWK/ LT, $\Delta$HEWK/$\Delta\tau$ > 0 ; HEWK/ LT $\Delta$HEWK/$\Delta\tau$ < 0

HEWW(htech) Global cumulative installed capacity in GW

HEWK(reg,htech) Heating capacity in GW *(fi23)*

LT(htech) Technology lifetimes *{fx12}*

Marginal cost of heat generation (weighted average of the levelised cost)

(fi34) HSTC = sum(HGC1*HEWG)/RHUD

HSTC(reg) Marginal cost of heat generation (\$/kWh)

HEWG(reg,htech) Heat generation by technology *(fi22)*

HGC1(reg,htech) Levelised cost for generating heat incl. policies, \$/kWh *(fi24)*

RHUD(reg) Total heat generation, GWh/y {fx17}

FTT:Heat policy inputs and exogenous values

{fx12} BHTC(reg,htech): Technology characteristics (capital, fuel, O&M, lifetimes, load factors etc)

{fx13} HTVS(reg,htech): Capital cost subsidies (in % paid by the gov.)

{fx14} HEFI(reg,htech): Feed-in tariffs (\$/kWh of generated heat)

{fx15} HREG(reg,htech): Regulations (whether a technology can be built or not, maximum share

limits)

{fx16} HTRT(reg): Fuel tax ($/kWh of energy content)

{fx17} RHUD(reg): Demand for heating services, GWh useful heat/y

### 1.2. Natural resource database and depletion algorithm

[f6] Fossil fuel depletion algorithm (non-linear differential equation)
NOTE: This is solved iteratively until ERRY and MERC become stable, solving for MERC: ERRY is global fossil fuel use, MERC is a marginal cost scaling global fossil fuel prices.

sum(ERRY) = sum(MPTR*BCSC(t)*H(C – MERC)$\Delta$C)
$\Delta$BCSC(t) = - MPTR*BCSC(t) $\Delta$t
NOTE: the sum is a discrete integral carried out over the cost axis C. This equation is evaluated iteratively until MERC and ERRY become stable.

| | |
|---|---|
| ERRY(reg,ft) | Global supply (=demand) of fossil fuels (k-toe/y) *(fi35)* |
| MPTR(reg,ft) | Reserve to production ratios (year$^{-1}$) *{fx18}* |
| BCSC(reg,ft,t,C) | Resources of fossil fuels left in the ground *{fx19}* |
| MERC(reg,eres) | Marginal cost of production ($/toe) *(fi36)* |
| C | Production cost independent temporary variable |

(fi35) Regional fossil fuel use (summed across fuel users)

ERRY = sum(FRCT), sum(FROT+FR04+FR05), sum(FRGT)

| | |
|---|---|
| ERRY(reg,ft) | Global supply (=demand) of fossil fuels (k-toe/y) |
| FRCT(reg,FU) | Coal use by fuel user *[2]* |
| FROT(reg,FU) | Heavy oil use by fuel user *[2]* |
| FR04(reg,FU) | Crude oil use by fuel user *[2]* |
| FR05(reg,FU) | Middle distillates (petrol, diesel, jet fuel, kerosene) by fuel user *[2]* |
| FRGT(reg,FU) | Natural gas use by fuel user *[2]* |

(fi36) Renewable energy resources use algorithm (cost-supply curves)

MERC = F(MEWG, BCSC)

| | |
|---|---|
| MERC(reg,eres) | Marginal cost of production ($/toe) |
| MEWG(reg,ptech) | Electricity generation in GWh/y *(fi3)* |
| BCSC(reg,ft,t,C) | Resources of fossil fuels left in the ground *{fx13}* |

FTT:Transport policy inputs and exogenous values

{fx18} MPTR Reserve to production ratios (year$^{-1}$)

{fx19} BCSC Database for non-renewable and renewable resources in 59 regions